\documentclass[preprint,review,12pt]{elsarticle} 

\usepackage{amssymb}
\usepackage{amsmath}
\usepackage{graphicx}
\usepackage{wrapfig}
\usepackage{lscape}
\usepackage{rotating}
\usepackage{epstopdf}
\usepackage{subfig}
\usepackage{lipsum}
\usepackage{url}
\usepackage{xcolor}

\journal{JQSRT}

\begin{document}
\begin{frontmatter}

\title{The VIIRS-DNB radiance product is insufficient to assess the effect of ``cool pavement'' materials on nighttime radiances of treated areas}

\author{John C.~Barentine} 
\affiliation{organization={Dark Sky Consulting, LLC},
            addressline={PMB 237, 9420 E. Golf Links Rd. Ste 108}, 
            city={Tucson},
            postcode={85730-1317}, 
            state={AZ},
            country={USA}}
\affiliation{organization={DarkSky Southern Arizona},
            addressline={5049 East Broadway Blvd, Suite 105}, 
            city={Tucson},
            postcode={85711-3646}, 
            state={AZ},
            country={USA}}

\begin{abstract}
Under increasing stress due to climate change, cities are looking for ways to mitigate its impacts. In warmer climates, they face the prospect of higher air temperatures in summer compared to historical averages due to the urban heat island effect. An approach intended to address this problem is the application of “cool pavement” treatments (CPT) to city streets to make them more reflective to sunlight. CPT raise the albedo of road surface materials such as asphalt concrete, causing them to absorb less energy from direct insolation and therefore radiate less heat at night. Raising the albedo of roadways for this purpose may also have the effect of increasing the amount of street light that is reflected into the night sky. The simplest hypothesis explaining the relationship between CPT application and upward radiance is that CPT applied to road surfaces in areas where street lighting is dominant should increase the upward radiance of neighborhoods where the treatments are applied. A simple model predicted radiance increases of 2-6\% immediately after CPT application. To test the hypothesis and model predictions, we looked for radiance changes coinciding with the application of CPT in residential neighborhoods of Phoenix, U.S., since 2020. We obtained time series radiance measurements from Visible Infrared Imaging Radiometer Suite Day-Night Band (VIIRS-DNB) data from Radiance Light Trends for Phoenix neighborhoods receiving CPT and nearby “control” neighborhoods not receiving CPT. At the 95\% confidence level, we found that any increases in nighttime radiances from treated neighborhoods did not exceed about 14\%. As a consequence, we cannot rule out either the expected radiance increases from our model or the possibility that CPT application yielded no change in radiance. We therefore cannot draw robust conclusions about the potential influence of CPT deployment on skyglow given the limitations of the DNB as a data source.
\end{abstract}



\begin{keyword}
Light pollution \sep Radiance \sep Cool pavement \sep Remote sensing
\end{keyword}

\end{frontmatter}

%
%
\section{Introduction}
\label{sec1:intro}

\subsection{The environmental context}
Our planet continues to warm, and human activity is principally driving the change.~\cite{Asibey2023,IPCC2023} Global climate change has many known and suspected effects on both the environment and human beings, including human society,~\cite{Dietz2020} cultural heritage,~\cite{Sesana2021}, the transmission of infectious diseases~\cite{VanDeVuurst2023}, and mental wellbeing.~\cite{Crane2022} Even as various groups work to reduce the influences contributing to climate change, considerable research has been conducted to determine optimal means of mitigating its effects.~\cite{Fawzy2020} 

There is particular concern about how cities will fare in a warmer future.~\cite{Masson2020} In particular, it is well known that urban environments retain more daytime heat during overnight hours, and hence remain warmer at night, than rural sites. This is understood to result from the `urban heat island' (UHI) effect.~\cite{Jabbar2023}. In the UHI scenario, elements of the built environment such as concrete, asphalt and tarmac absorb solar radiation during the daytime and radiate it as heat overnight. In a warming climate, reducing the UHI effect is an urgent matter of public health.~\cite{Yadav2023} Higher overnight temperatures in cities lead to hotter following days, increasing the risk of morbidity and mortality due to heat stroke and other heat-related illnesses.~\cite{VicedoCabrera2021}

\subsection{Mitigating urban heat risk by modifying the built environment}
Researchers and engineers have suggested a number of ways of mitigating heat risk in cities, including approaches like bolstering green infrastructure~\cite{Balany2020} and passive daytime radiative cooling.~\cite{Kousis2023a} Raising the albedo of surfaces in the built environment to reduce the amount of absorbed insolation during the day in turn reduces the amount of heat they radiate at night. Given the prevalence of impervious cover in many cities, attention has recently turned to manipulation of the radiative properties of those materials.~\cite{Vujovic2021,Wardeh2022} Higher thermal emissivity of surfaces leads to lower surface temperatures. ``Cool pavement'' treatments (CPT) aim to modify pavement surfaces by overlaying materials on top of them that change their reflective, evaporative and thermal properties.

Materials research and field experiments demonstrate that CPT have significant air temperature reduction potential in the urban environment, both during the day and at night.~\cite{Kappou2022,Seifeddine2023} A number of factors influence the temperature of the air near pavements, including their albedo, thermal conductivity, heat capacity and emissivity. Among these, albedo dominates.~\cite{Qin2022} The use of CPT therefore aims to raise the albedo of the pavements they cover. In this way, CPT exceeds the cooling capability of other approaches by as much as 25$^{\circ}$C.~\cite{Kousis2023b} For this reason, CPT has been suggested as a means of lowering temperatures in cities specifically to benefit public health.~\cite{Cooley2024} At the same time, a challenge remains in finding constituent CPT materials that are durable in the environment and adhere well to surfaces, and yet also do not yield pollution.~\cite{Wang2022}

Many of the CPT materials tested to date appear to have high albedo in the optical as well as the infrared. To the extent that surfaces like roadways are illuminated at night by overhead lighting, they are diffuse reflectors of optical-wavelength light. This can increase the glare associated with roadways at night.~\cite{Anupam2020} It is understood that street lighting accounts for much of the total nighttime light emissions in many world cities.~\cite{Luginbuhl2009,Kuechly2012,Hiscocks2010,Bara2018,Kyba2019} Anecdotally, in low-density U.S. residential neighborhoods, street lighting dominates overnight light emissions as there are few other sources active from dusk to dawn each night. Raising the albedo of the pavements that street lighting illuminates may therefore increase the amount of artificial light reflected into the night sky, which may manifest as an increase in the upward-directed radiance of cities as seen from remote sensing platforms such as Earth-orbiting satellites.

\subsection{Motivation for this study}
Our concern in this particular study has to do with the potential for increased upward-directed radiance from CPT reflections to increase the brightness of the night sky in and near cities. Skyglow has immediate implications for the visual quality of the night sky~\cite{Varela2023} and the observability of celestial phenomena that can hamper astronomical discovery.~\cite{Green2022,Falchi2022} It can also raise the total illuminance at ground level,~\cite{Bara2022} which has ecological consequences.~\cite{Sanders2020,Brayley2022} Efforts to reduce the environmental and social harm associated with one influence (climate change) therefore might well induce harm due to another (light pollution). We therefore sought to determine whether there were any such concerns stemming from the potential large-scale deployment of CPT by cities as a climate change resilience measure.

There is little information about this particular topic in the literature to date, and most of that discussion is in the context of lighting research. Rossi, Iacomussi and Zinzi~\cite{Rossi2018} reported field and laboratory measurements of two kinds of elastomeric pavement coatings applied to worn asphaltic surfaces. They found that the spectral reflectivities of the treated surfaces at optical wavelengths were between about four and 15 times higher than those of the untreated surfaces. They also observed strong Lambertian diffusing behaviors among the treated surfaces. On this basis, they argued that the installed power of roadway lighting systems in areas where CPT are deployed should be reduced in order to limit excessive upward light emissions expected to otherwise contribute to increased skyglow. Similar conclusions were reached in other photometric studies of pavement materials not immediately relevant to CPT.~\cite{Gidlund2019,Muzet2022}

\subsection{Hypothesis}
This leads to the question considered here: Do cool pavement treatments that raise the albedo of roadways to reduce the urban heat island effect also increase the diffuse reflection of street lighting into the night sky? The visual appearance of freshly treated road surfaces compared to the materials they cover (generally, asphalt and tarmac) suggests high albedo at optical wavelengths. The ``naive'' hypothesis is that CPT applied to worn road surfaces will temporarily cause enhanced reflectivity, but this effect will quickly fade as the treatment material is exposed to weathering. That is, a breakdown or wearing away of CPT material will tend to reveal the darker substances, reducing albedo. Depending on the durability of the CPT material and the extent to which it is degraded by environmental exposure, any change in the upward radiance may well be short-lived.

A simple model of the situation compares the measured change in upward radiance $L$ at the top of the Earth's atmosphere after the application of CPT materials, ${\Delta}L=L_\textrm{after}-L_\textrm{before}$, to the radiance before application, $L_\textrm{before}$. The radiance depends on both the change in the reflectivity $r$ of asphalt concrete pavements before and after application of CPT materials, ${\Delta}r=r_\textrm{after}-r_\textrm{before}$,  and the fraction $a$ of the territory over which the application takes place:
\begin{equation}
\frac{{\Delta}L}{L_\textrm{before}} = a\left[\frac{r_\textrm{after}}{r_\textrm{before}}-1\right].
\end{equation}
For measured values of ${r_\textrm{after}}/{r_\textrm{before}}$ of about 1.5 and typical values of $a$ around 0.06 (see Section~\ref{subsec:predictions}), this model predicts radiance changes of about +3\%. This study is in part concerned with the ability to detect changes of this order in nighttime radiance time series obtained by Earth-orbiting satellites.

The remainder of this paper is organized as follows. In Section~\ref{sec:method} we describe the experimental approach in detail. Results are analyzed and interpreted in Section~\ref{sec:analysis}. Finally, we summarize our findings and the conclusions we draw from them in Section~\ref{sec:summary}.

%
%
\section{Experimental method}
\label{sec:method}
Testing the hypothesis stated previously is conceptually straightforward. Observing the upward radiance of treated areas at night before and after CPT is applied should tend to reveal any increase in brightness with respect to radiances of nearby `control' areas not receiving the treatment. Long time series of observations are useful in characterizing measurement uncertainties so as to evaluate the statistical significance of any apparent increase in radiance possibly attributable to the deployment of CPT. 

An ideal location to test the hypothesis would have several characteristics. First, it would be a generally arid environment with frequently clear weather and typically low aerosol optical depth. A dense urban location would ensure an extensive roadway network to sample a sufficient length of both treated and untreated control surfaces. Finally, it would be lit extensively by roadway lighting at least during the times of night during which the radiance time series was sampled. 

As our test city we chose Phoenix, Arizona, U.S. It has a dry climate; clear weather prevails for much of the year and roadways in the city are typically dry. There is no snow or ice on the ground during the winter months that tends to amplify surface light reflections.~\cite{Levin2017,Jechow2019} Within the incorporated boundaries of the City of Phoenix there are over 7800 km$^2$ of paved public streets. Residential neighborhoods are illuminated during the overnight hours by fully shielded, white light-emitting diode (LED) luminaires whose nominal correlated color temperature is 2700 K.

Among various possible heat sources, pavement infrastructure contributes significantly to Phoenix’s urban heat balance,~\cite{Hoehne2020} which makes it an attractive target for insolation reduction through the use of reflective overcoating. Phoenix began deploying CPT on an experimental basis in 2019 to address both the city's long-term sustainability and rising heat-related illness and death among its residents during the summer months.~\cite{Watkins2021,duBray2023} It applied CoolSeal, a CPT material manufactured by GuardTop LLC of Laguna Niguel, California, U.S. The material, consisting of asphalt, water, an emulsifying agent, mineral fillers and polymers, is applied on top of existing asphalt pavements. Field data collected after CPT deployment shows decreased temperatures in line with expectations.~\cite{Elmagri2024} 

\begin{figure}[ht]
\centering
\includegraphics[width=.95\linewidth]{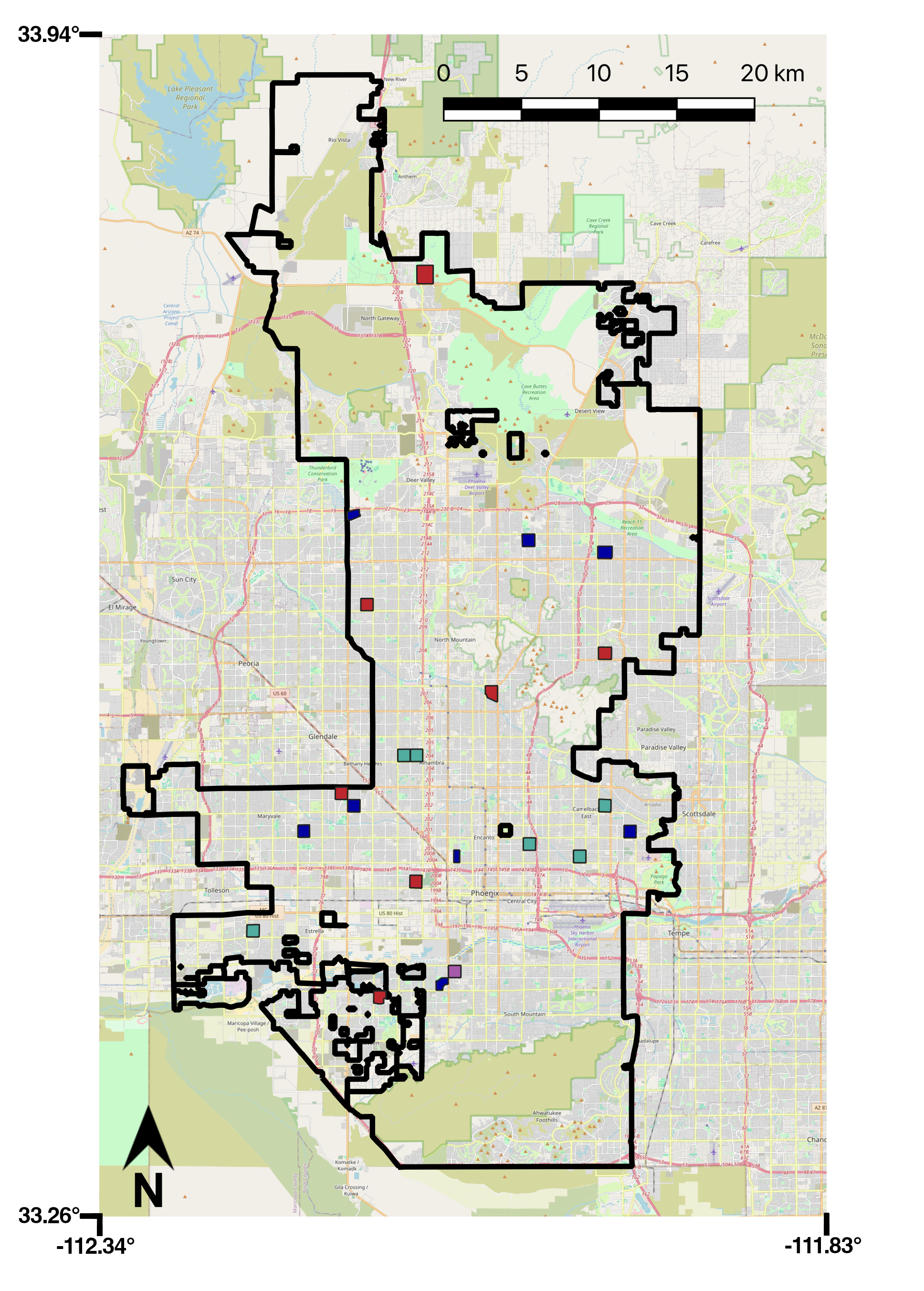}
\caption{Map showing the locations of Phoenix, Arizona, residential neighborhoods receiving CPT between 2020 and 2023. The incorporated boundaries of the City of Phoenix are indicated by the heavy solid line. Treated areas are represented by solid polygons whose colors correspond to the year in which they were treated: 2020 (red), 2021 (purple), 2022 (blue) and 2023 (teal). The basemap is from OpenStreetMap®, open data licensed under the Open Data Commons Open Database License (ODbL) by the OpenStreetMap Foundation (OSMF).}\label{CPT-map}
\end{figure}

We adopted the same identifying nomenclature used by the City of Phoenix to designate areas receiving CPT. This is compiled in Table~\ref{CPT-summary} along with basic information about each treated area: the coordinates of its geographic center, the treatment start and end dates, and the total length of roadway surface receiving the treatment. The locations of the treated areas are shown in Figure~\ref{CPT-map}.
\begin{table}
\centering
\begin{tabular}{ lccccc }
 \hline
\textbf{} & \textbf{Center} & \textbf{Center} & \textbf{Treatment} & \textbf{Treatment} & \textbf{Treatment} \\
\textbf{Code} & \textbf{Lat.~($^{\circ})$} & \textbf{Long.~($^{\circ})$} & \textbf{Start Date} & \textbf{End Date} & \textbf{Length (km)} \\ 
\hline
QS 33-18 & 33.61436 & -112.15550 & 2020.667 & 2020.675 & 5.63 \\
QS 59-23 & 33.80389 & -112.11539 & 2020.828 & 2020.850 & 11.23 \\
QS 29-37 & 33.58639 & -111.99114 & 2020.725 & 2020.783 & 8.92 \\
QS 11-22 & 33.45495 & -112.12169 & 2020.631 & 2020.650 & 8.47 \\
QS 18-16 & 33.50580 & -112.17313 & 2020.653 & 2020.661 & 4.20 \\
QS 26-28 & 33.56375 & -112.06918 & 2020.822 & 2020.842 & 7.84 \\
QS 2-19 & 33.38849 & -112.14686 & 2020.794 & 2020.814 & 6.50 \\
QS 4-25 & 33.40314 & -112.09482 & 2021.800 & 2021.811 & 5.71 \\
QS 40-17 & 33.66604 & -112.16496 & 2022.339 & 2022.358 & 4.89 \\
QS 37-37 & 33.64437 & -111.99113 & 2022.361 & 2022.383 & 9.32 \\
QS 38-31 & 33.65132 & -112.04395 & 2022.386 & 2022.403 & 6.14 \\
QS 13-25 & 33.46951 & -112.09353 & 2022.872 & 2022.875 & 3.21 \\
QS 17-17 & 33.49858 & -112.16453 & 2022.461 & 2022.492 & 9.08 \\
QS 15-39 & 33.66604 & -112.16496 & 2022.394 & 2022.397 & 5.87 \\
QS 15-13 & 33.48401 & -112.19916 & 2022.417 & 2022.433 & 6.43 \\
QS 3-24 & 33.39615 & -112.10395 & 2022.442 & 2022.458 & 7.22 \\
QS 14-31 & 33.47657 & -112.04328 & 2023.331 & 2023.406 & 4.81 \\
QS 21-22 & 33.52764 & -112.12125 & 2023.278 & 2023.289 & 6.08 \\
QS 21-21/22 & 33.52761 & -112.12994 & 2023.297 & 2023.308 & 5.95 \\
QS 17-37 & 33.49868 & -111.99116 & 2023.400 & 2023.631 & 6.11 \\
QS 7-9 & 33.42657 & -112.23416 & 2023.422 & 2023.683 & 7.81 \\
QS 13-35 & 33.46942 & -112.00858 & 2023.317 & 2023.336 & 8.15 \\
\end{tabular}
\caption{Basic data for each of the Phoenix, Arizona, neighborhoods receiving CPT between 2020 and 2023.}
\label{CPT-summary}
\end{table}

We obtained time series of nighttime upward radiance measurements before and after CPT was applied in Phoenix neighborhoods and compared them to time series measurements from similar untreated areas nearby. The source of the radiance measurements was the Visible Infrared Imaging Radiometer Suite (VIIRS) aboard the Suomi National Polar-orbiting Partnership satellite. In particular, we used measurements made in the VIIRS Day-Night Band (DNB).~\cite{Lee2006,Liao2013} The DNB is is capable of imaging nighttime terrestrial light emissions at wavelengths between 500 and 900 nm to a nominal radiance limit of 3 nW cm$^{-2}$ sr$^{-1}$. Its data have been used extensively to study the spatiotemporal dynamics of artificial light at night as seen from Earth orbit.~\cite{Levin2020} As a nighttime lights sensor the DNB has some drawbacks, such as its coarse spatial resolution (about 750 m pix$^{-1}$) and relative insensitivity to much of the light in the spectral power distribution (SPD) of white LED lighting sources. However, it is currently the only source of whole-Earth measurements of broadband upward radiance with a nightly cadence whose data are freely available to the public.

Radiance time series were extracted from DNB data using the Radiance Light Trends (RLT) web tool.~\cite{RLT} RLT is noted for its user-friendly interface and rapid generation of radiance time series for polygons of arbitrary shape up to 10000 km$^2$ in area. To make the time series it uses monthly average cloud-free composite radiances for each DNB pixel computed by the Earth Observation Group at the Colorado School of Mines Payne Institute for Public Policy.~\cite{Elvidge2017} Before radiances are averaged for the month, they are filtered to exclude images affected by stray light, lightning, lunar illumination and cloud cover. The data are also corrected for a zeropoint offset of 0.2 nW cm$^{-2}$ sr$^{-1}$.~\cite{Coesfeld2020} Our analysis was concerned only with searching for relative changes in a time series of radiances, so a consistent absolute calibration of all radiances was not important. Lastly, RLT reprojects the DNB pixels onto a geospatially consistent grid. 

\begin{figure}[ht]
\centering
\includegraphics[width=1.0\linewidth]{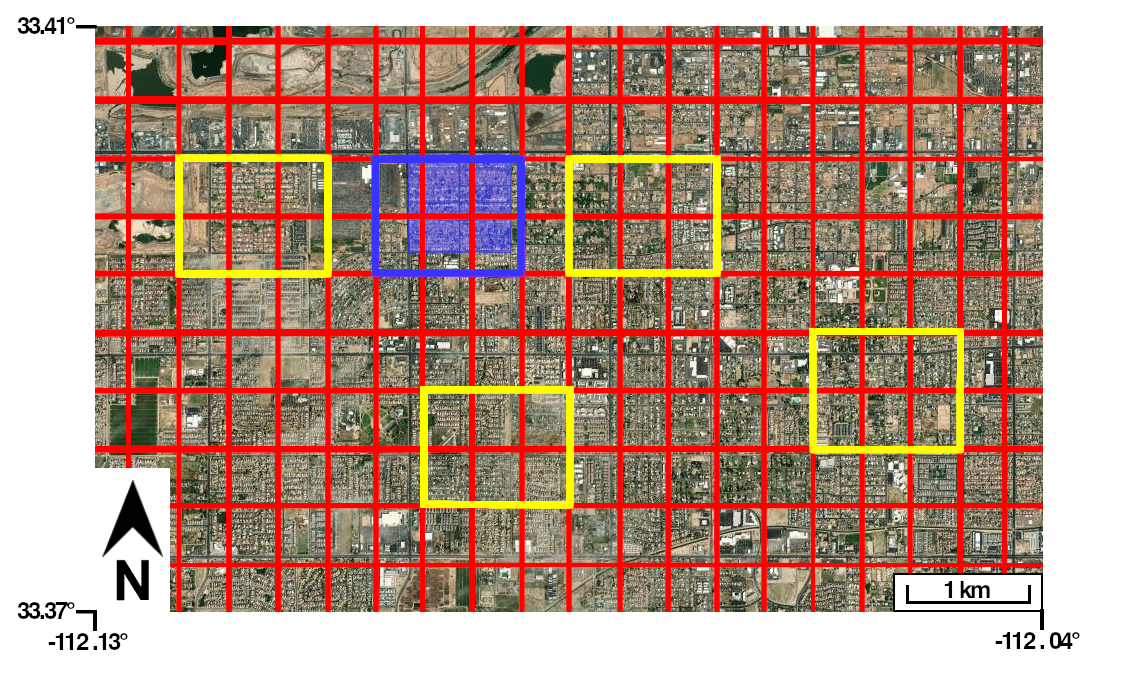}
\caption{An example of the selection of treated and comparison areas for radiance time series extraction. The background is an optical-wavelength satellite image of Phoenix obtained under daytime conditions. The red grid represents georeferenced DNB pixels projected onto this image. The group of 2$\times$3 pixels outlined in blue is the smallest cluster of contiguous DNB pixels fully containing the treated neighborhood designated ``QS 4-25'' by the City of Phoenix; the area receiving CPT is shaded in the same color. Yellow 2$\times$3-pixel rectangles mark the boundaries of ``control'' areas not receiving CPT but consisting of comparable land-use types and intensities.}\label{QS-4-25-selections}
\end{figure}

We obtained information from the City of Phoenix government about the residential neighborhoods receiving CPT in the Cool Pavement Program,\footnote{\url{https://www.phoenix.gov/streets/coolpavement/}} including location, the start and end dates of CPT application, and the number of miles of treated roadway surface. For each neighborhood treated with CPT, we defined sample and comparison areas consisting of the same whole number of contiguous DNB pixels. First we determined the smallest number of contiguous DNB pixels that fully enclosed the treated area. Next, we identified four nearby, untreated control areas in which the land use types and intensities were similar to that of the treated area (see Figure~\ref{QS-4-25-selections}). To avoid cross-contamination and limit the influence of instrumentally scattered light, we avoided pixel clusters as comparisons that were contiguous to the treated area along even a single pixel boundary. We extracted a time series of average radiances from each pixel cluster and corrected them for projection by multiplying by the cosine of the latitude of the geometric center of each cluster.

\begin{figure}
\centering
\includegraphics[width=1.0\linewidth]{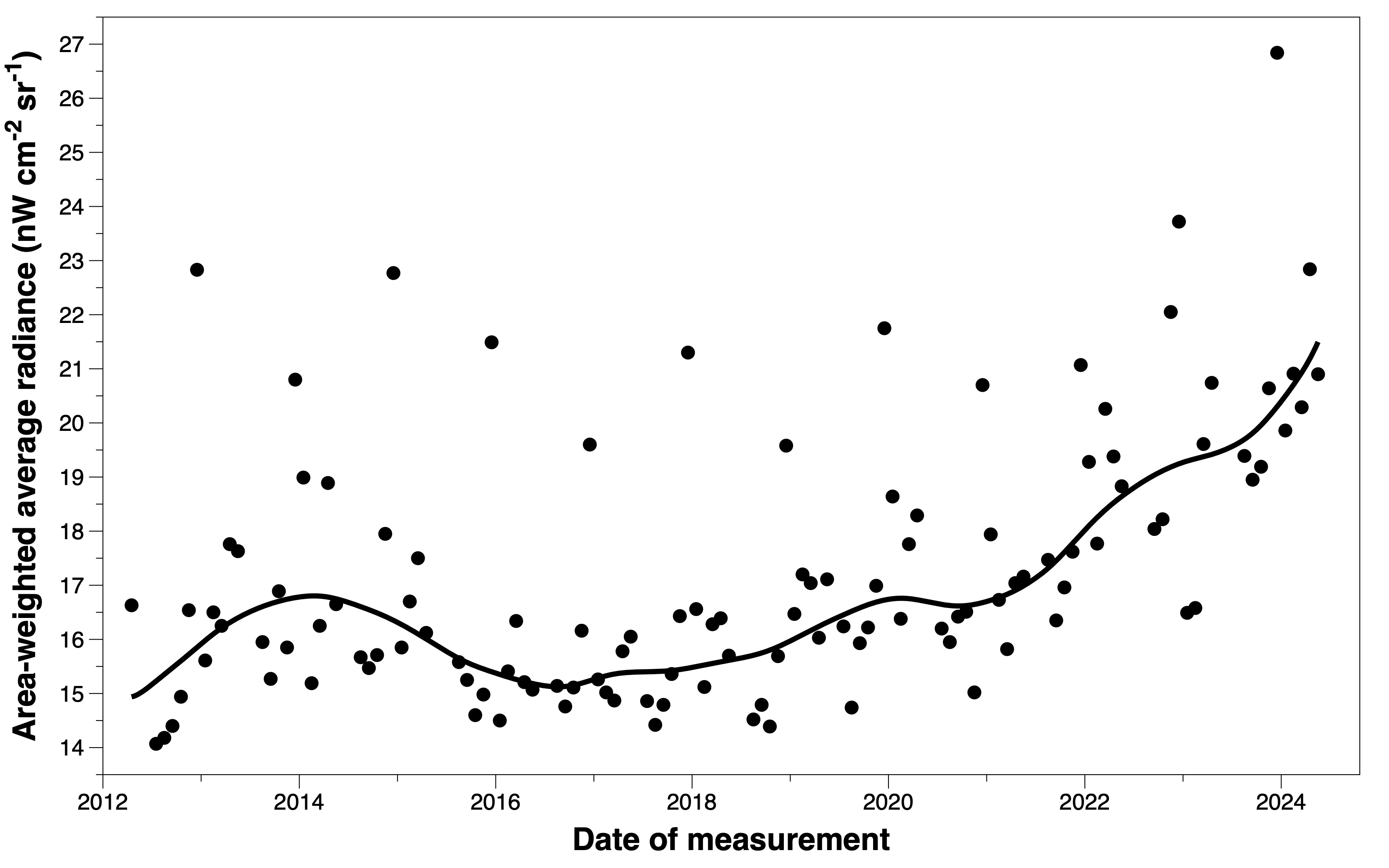}
\caption{An example of a LOESS fit to the DNB radiance time series data. The solid points are the radiances for the neighborhood designated QS 4-25 that received CPT in October 2021. The line shows the LOESS fit to the underlying, long-term trend. Outliers to the an initial fit with values ${\geq}2\sigma$ were rejected and the fit performed a second time.}
\label{LOESS-fit-example}
\end{figure}
To characterize the scatter in the time series radiances, we fitted both the treated and control areas using the locally estimated scatterplot smoothing (LOESS) method. LOESS is a non-parametric regression method that combines multiple regression models using a $k$-nearest neighbors algorithm-based meta-model.~\cite{Cleveland1988} 
The fits were trimmed of outliers beyond 2${\sigma}$ to get at the underlying variation in the radiances over long time periods. In particular, this removed distinct influences such as the tendency of December measurements to be anomalously high (see Section~\ref{sec:analysis}). An example of a LOESS fit to the radiance time series for a treated Phoenix neighborhood is shown in Figure~\ref{LOESS-fit-example}.

\begin{figure}[ht]
\centering
\includegraphics[width=1.0\linewidth]{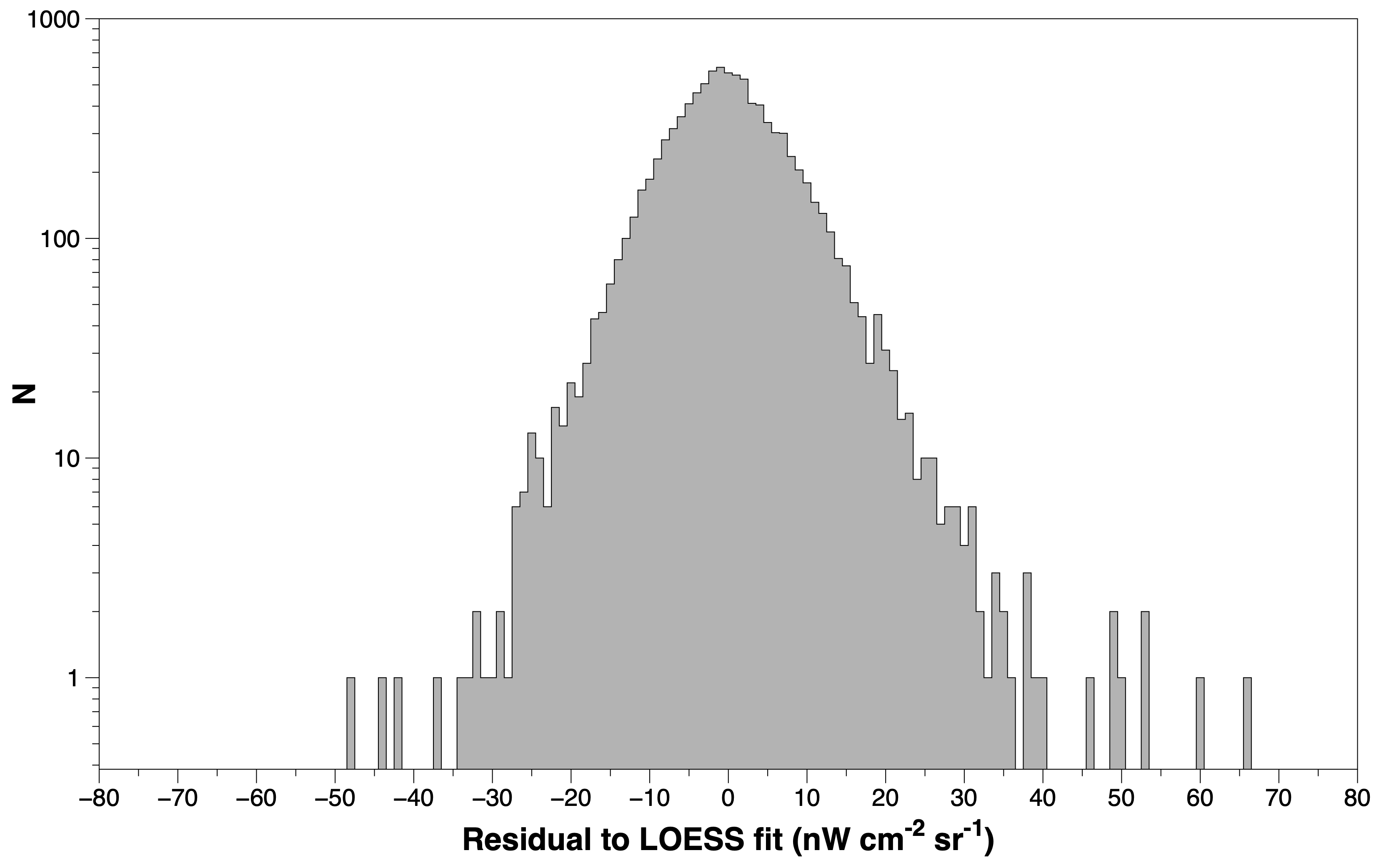}
\caption{Histogram of residuals to locally estimated scatterplot smoothing (LOESS) fits of secular trends in radiance time series for all sample areas in this study. The histogram is plotted with a log$_{10}$ scale on the ordinate to more clearly show the outliers. The data are well-fitted with a normal distribution whose mean and first standard deviation are 0.03 and 6 nW cm$^{-2}$ sr$^{-1}$, respectively.}
\label{LOESS-fit-residuals-hist}
\end{figure}
If the fits adequately remove the remaining, long-term variations, then the fit residuals should be normally distributed about zero. This can be seen in the histogram plot in Figure~\ref{LOESS-fit-residuals-hist}. To test the goodness of fit, we applied one-sample Anderson-Darling tests to the fit residuals for the treated areas and the controls as two aggregate groups. This confirmed the Gaussianity of the residuals to $p<0.01$ ($N$=9602). Because we found no meaningful difference in the distribution of residuals from the fits to the treated areas versus the controls, we took the standard deviation of the fit residuals of all sampled areas (6 nW cm$^{-2}$ sr$^{-1}$) as an estimate of the scatter in the radiance time series.

%
%
\section{Analysis and interpretation}
\label{sec:analysis}

%
%
\begin{sidewaysfigure*}[t!]
\centering
        \subfloat[]{%
            \includegraphics[width=.48\linewidth]{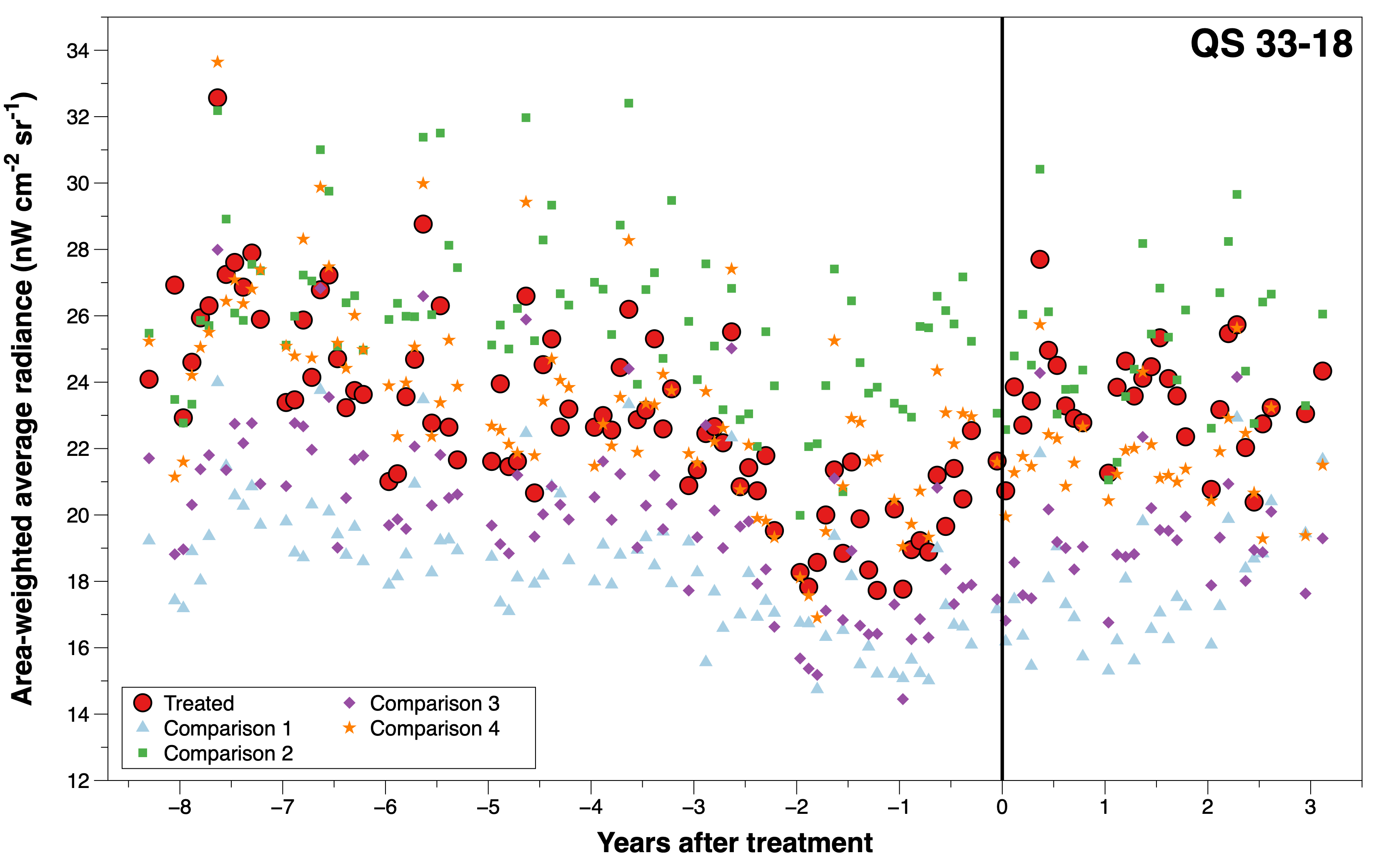}%
        }\hfill
        \subfloat[]{%
            \includegraphics[width=.48\linewidth]{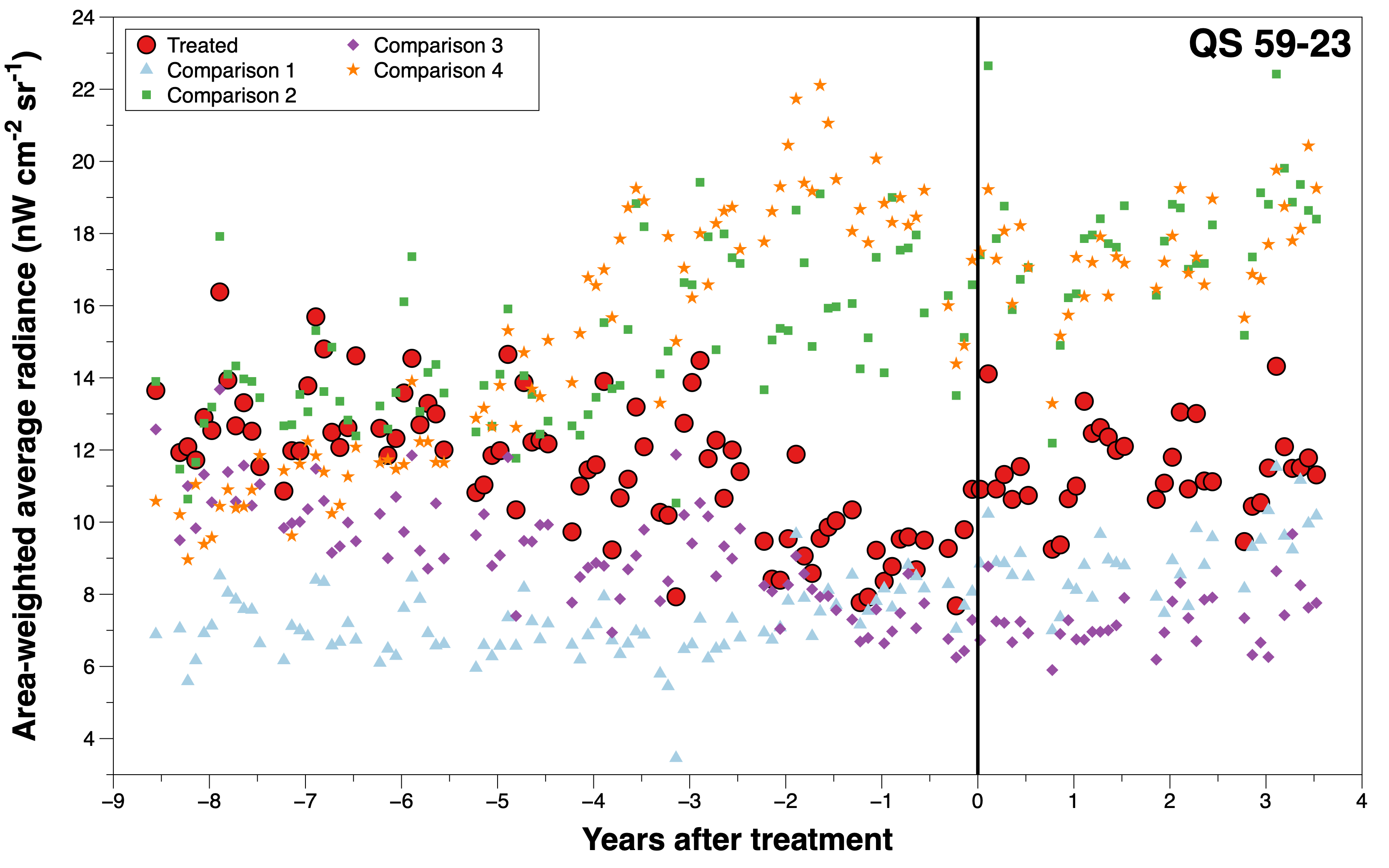}%
        }\\
        \subfloat[]{%
            \includegraphics[width=.48\linewidth]{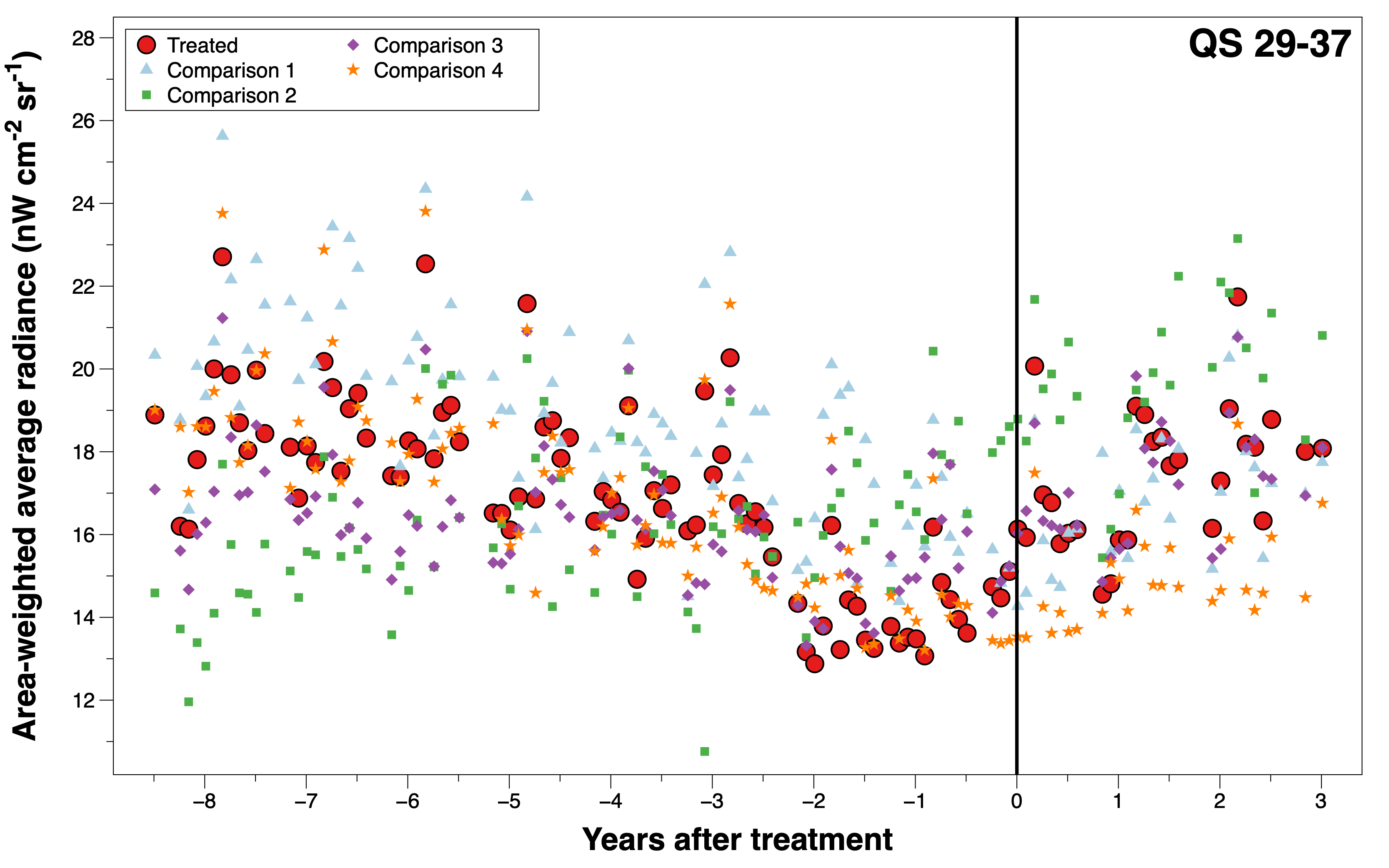}%
        }\hfill
        \subfloat[]{%
            \includegraphics[width=.48\linewidth]{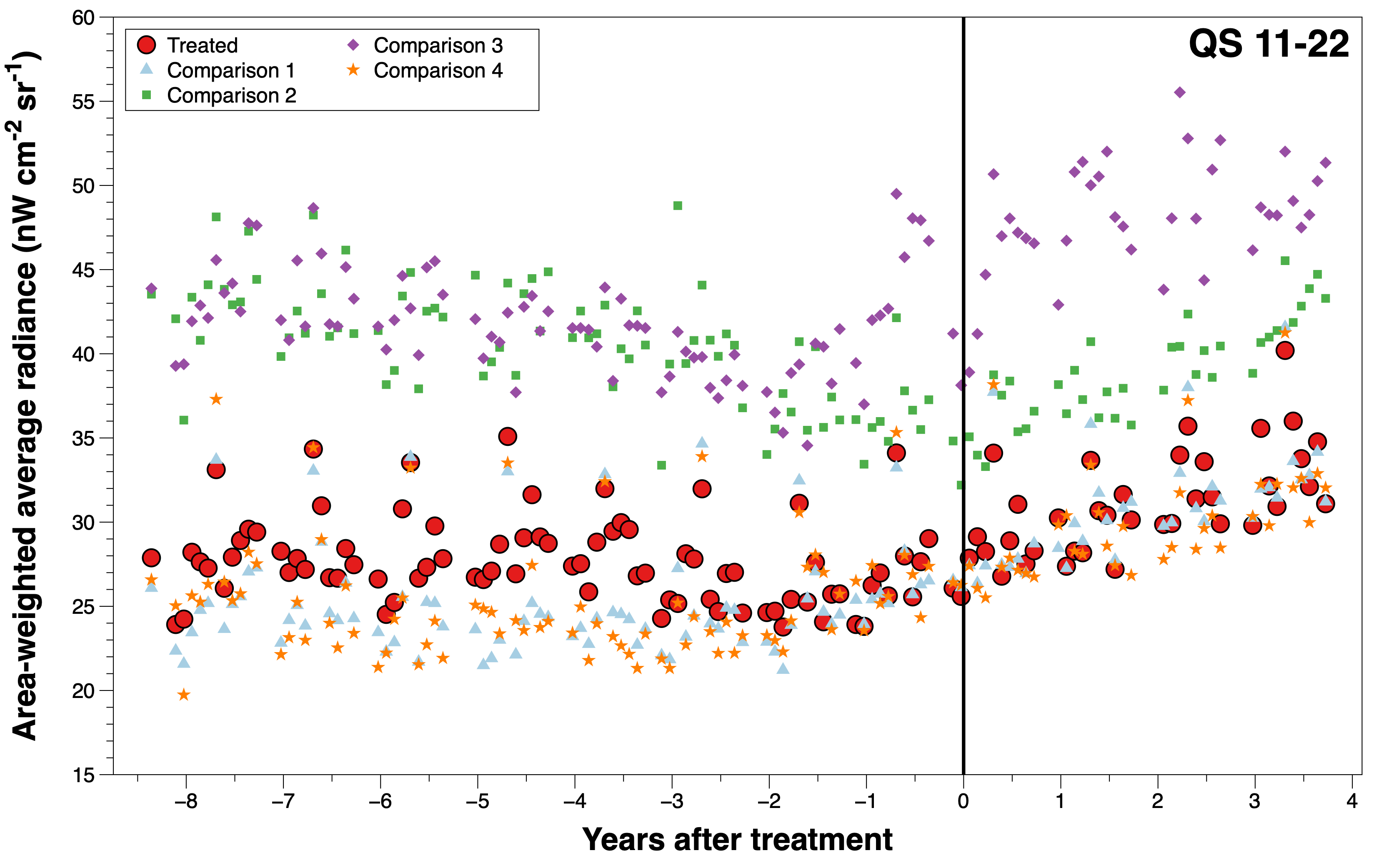}%
        }
        \caption{Upward radiance time series for Phoenix, Arizona, residential neighborhoods receiving CPT (`Treated') and control neighborhoods not receiving CPT (`Comparison' 1, 2, 3 and 4). The abscissa represents time in years relative to the application of CPT; a vertical line at $t=0$ is included to help guide the eye. Each panel is labeled in the upper right corner with the City of Phoenix code referring to the treated neighborhood shown; see Table~\ref{CPT-summary} for their locations and treatment dates.}
        \label{radiance-plots-1}
\end{sidewaysfigure*}

%
%
\begin{sidewaysfigure*}[t!]
\centering
        \subfloat[]{%
            \includegraphics[width=.48\linewidth]{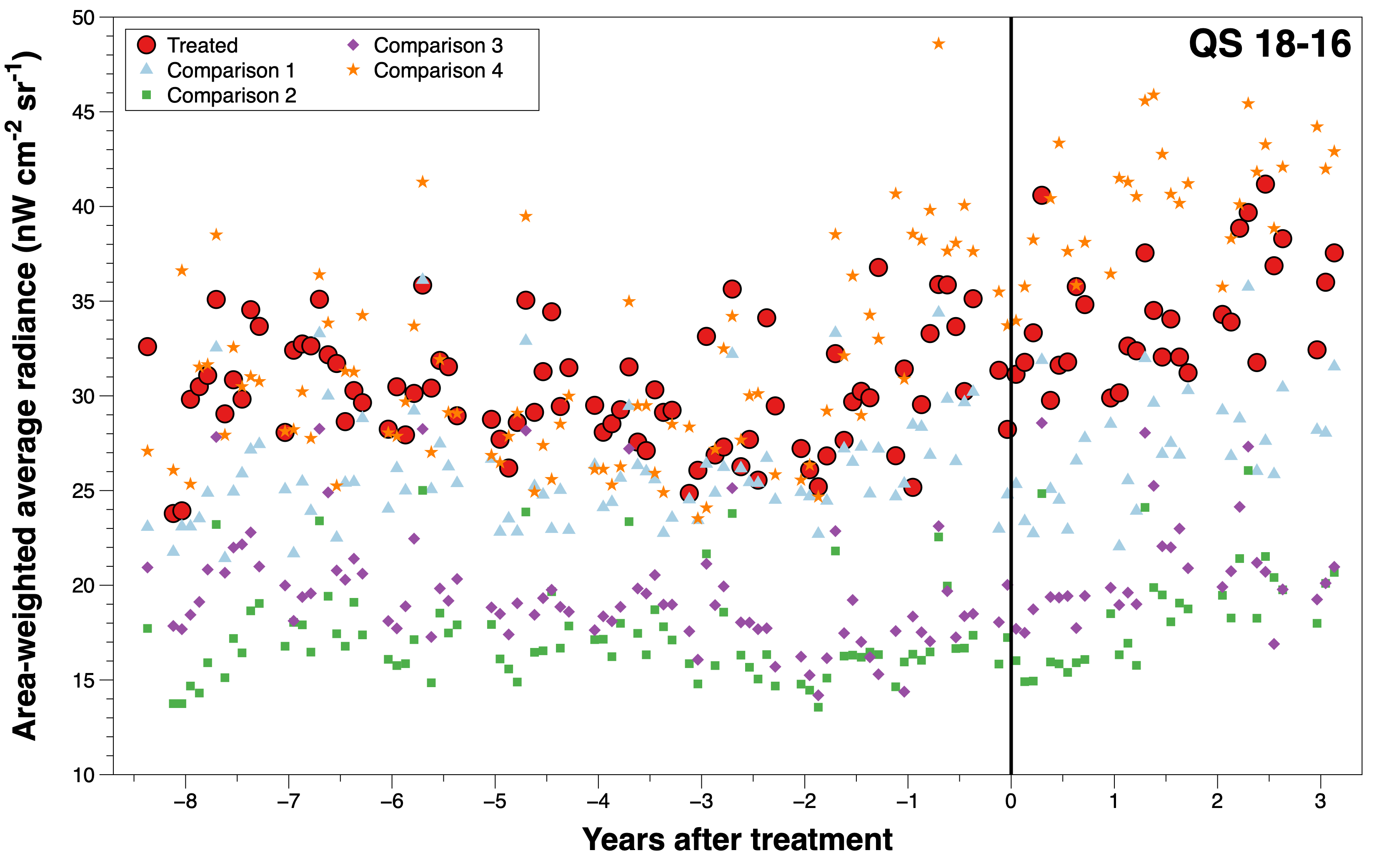}%
        }\hfill
        \subfloat[]{%
            \includegraphics[width=.48\linewidth]{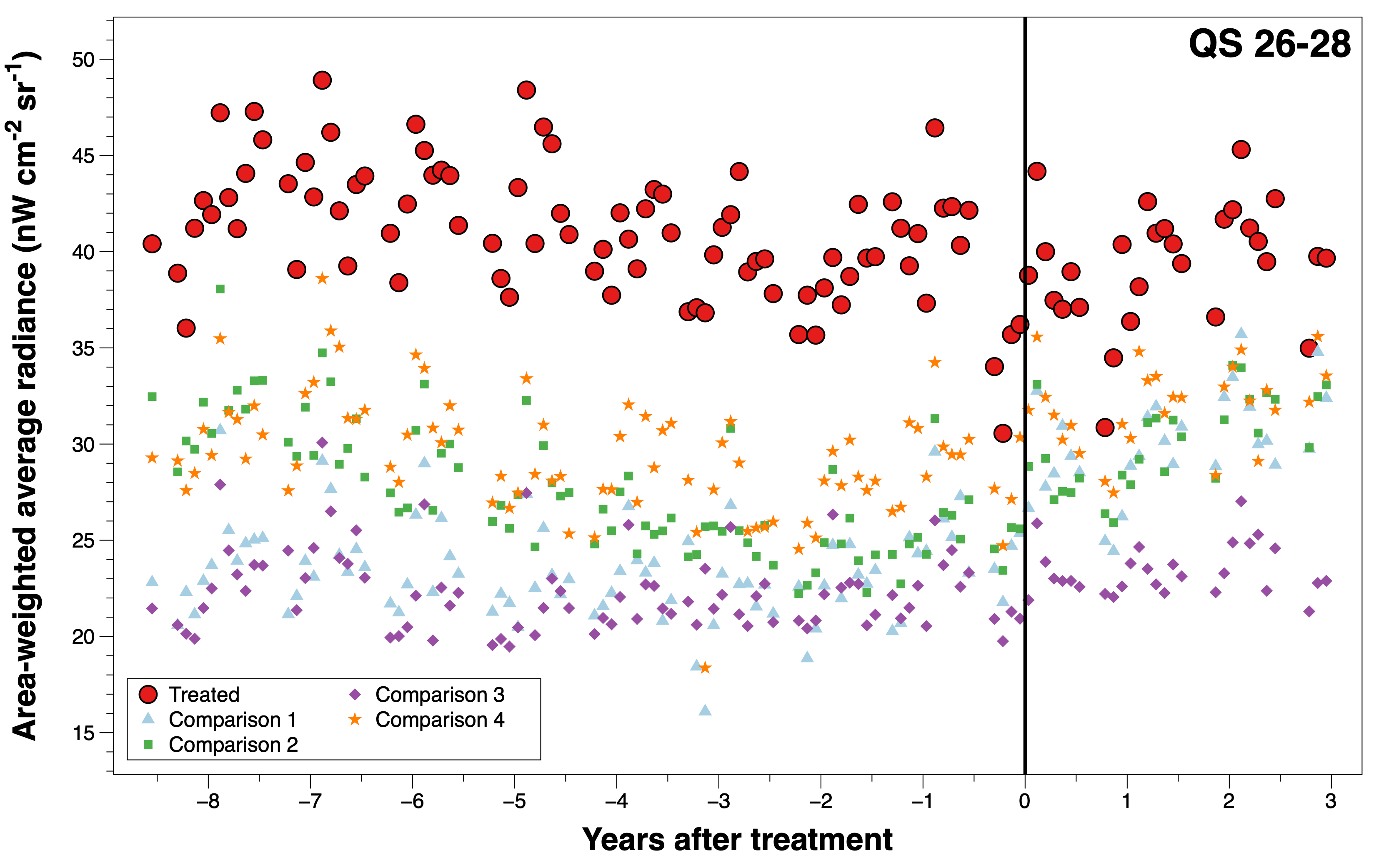}%
        }\\
        \subfloat[]{%
            \includegraphics[width=.48\linewidth]{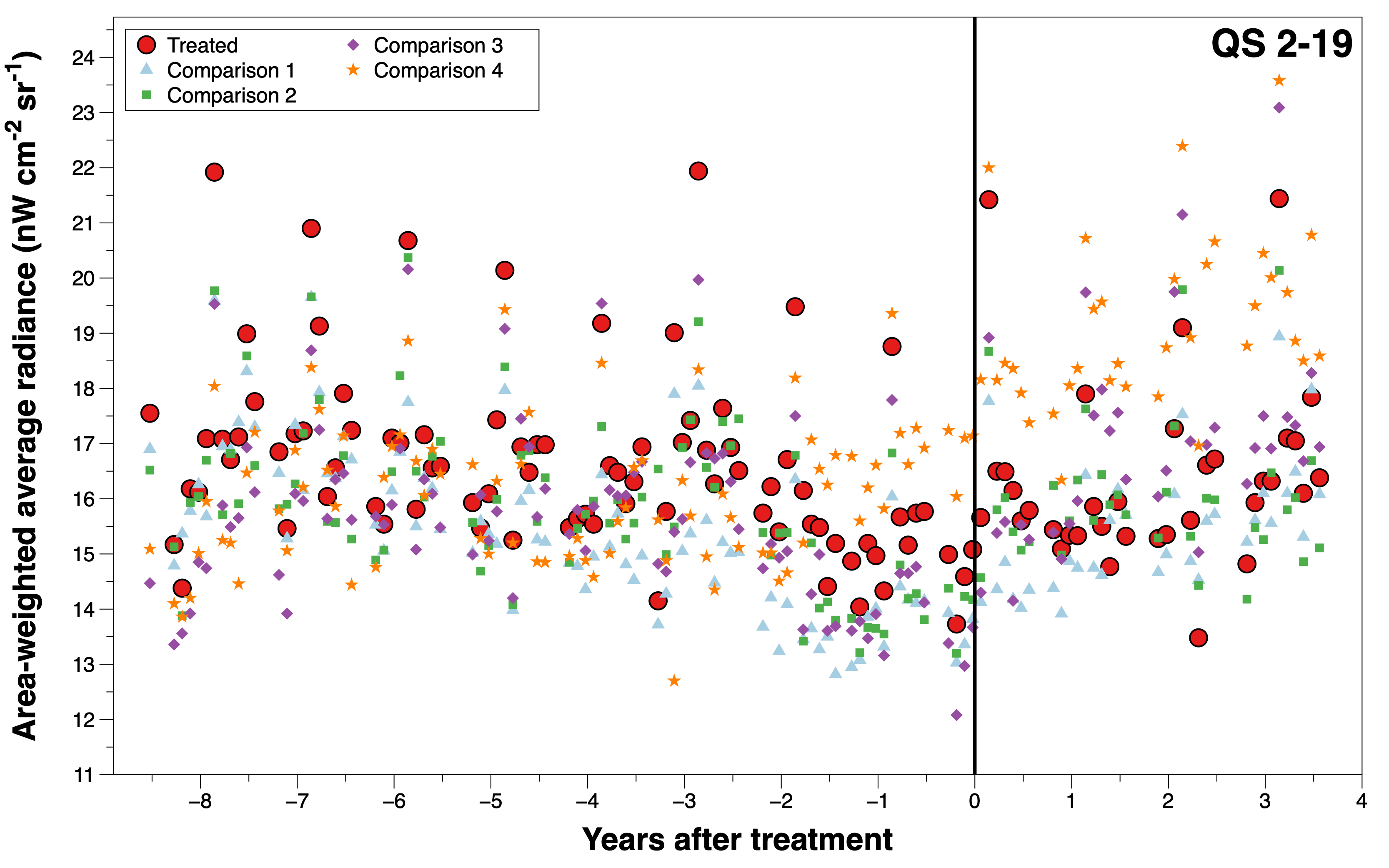}%
        }\hfill
        \subfloat[]{%
            \includegraphics[width=.48\linewidth]{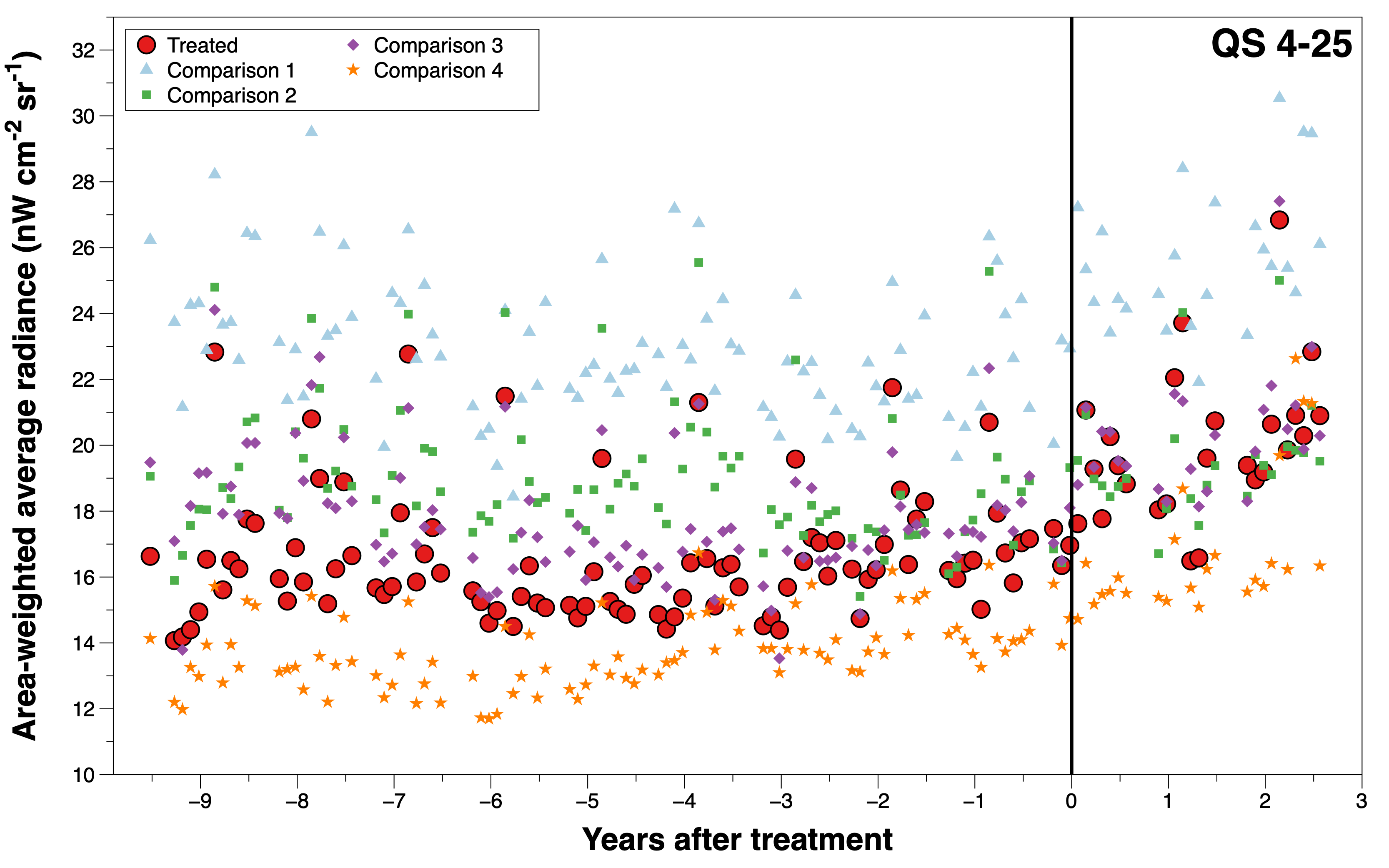}%
        }
        \caption{Upward radiance time series for Phoenix, Arizona, residential neighborhoods receiving CPT (`Treated') and control neighborhoods not receiving CPT (`Comparison' 1, 2, 3 and 4). Details are the same as in Figure~\ref{radiance-plots-1}.}
        \label{radiance-plots-2}
\end{sidewaysfigure*}

%
%
\begin{sidewaysfigure*}[t!]
\centering
        \subfloat[]{%
            \includegraphics[width=.48\linewidth]{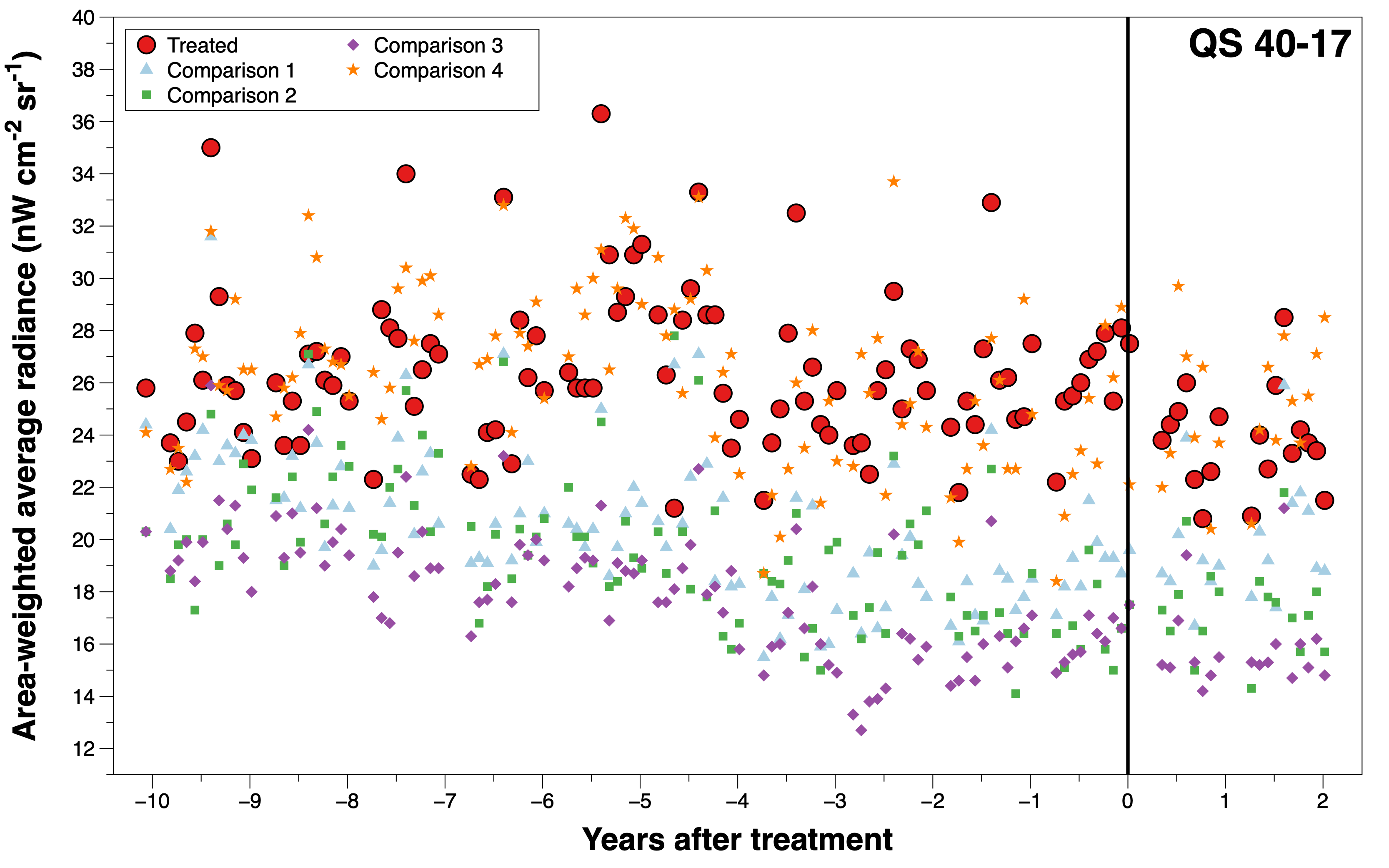}%
        }\hfill
        \subfloat[]{%
            \includegraphics[width=.48\linewidth]{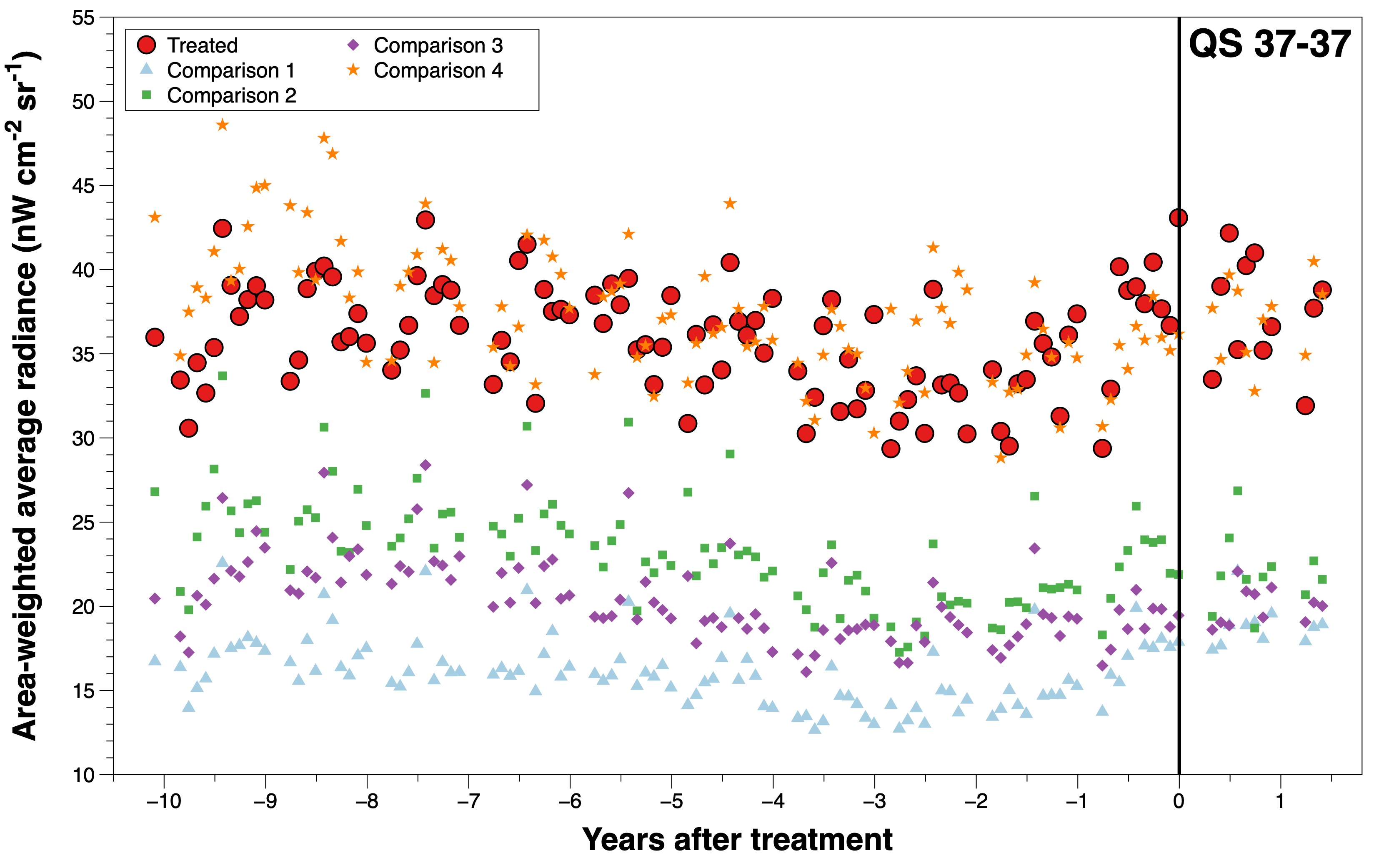}%
        }\\
        \subfloat[]{%
            \includegraphics[width=.48\linewidth]{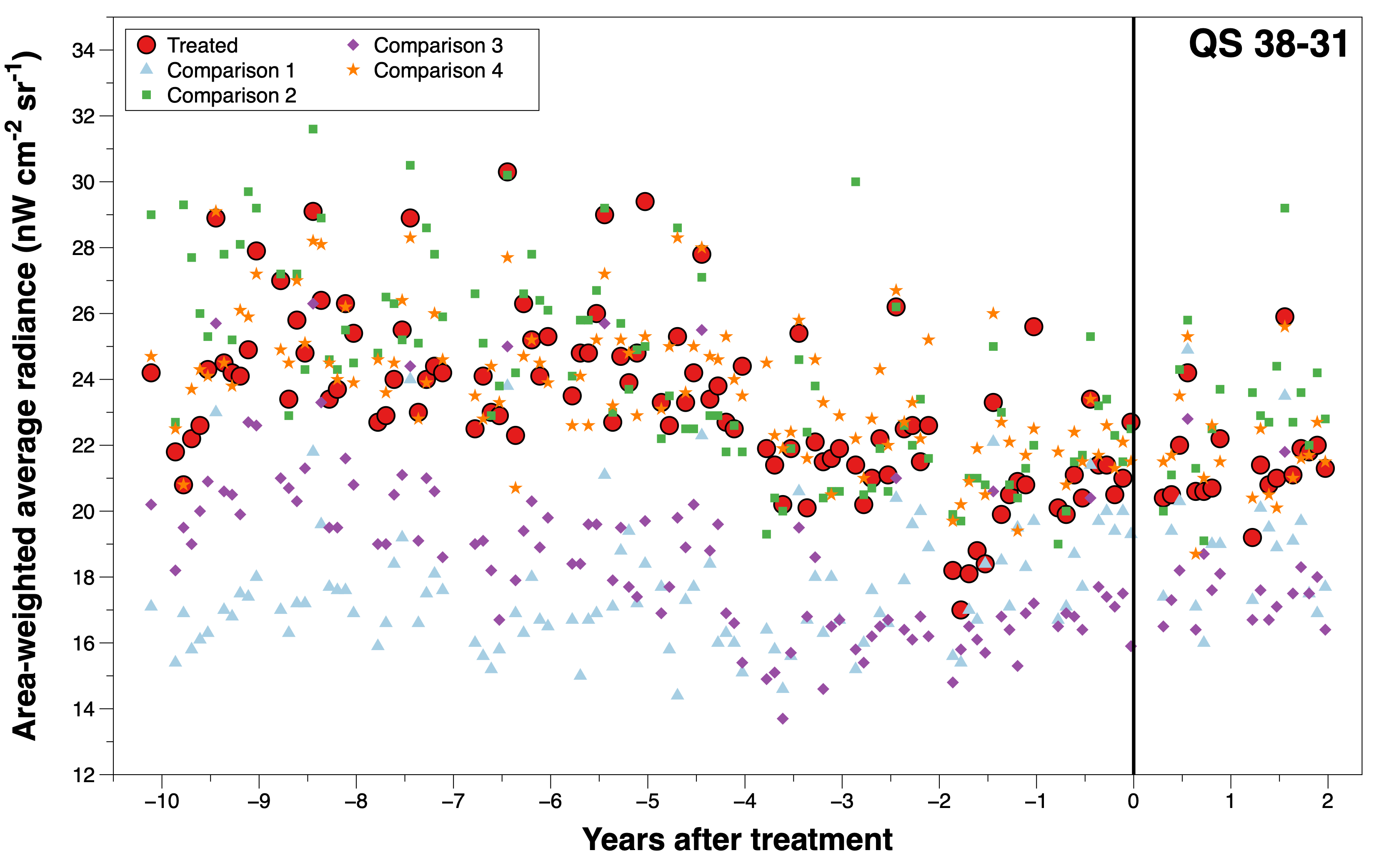}%
        }\hfill
        \subfloat[]{%
            \includegraphics[width=.48\linewidth]{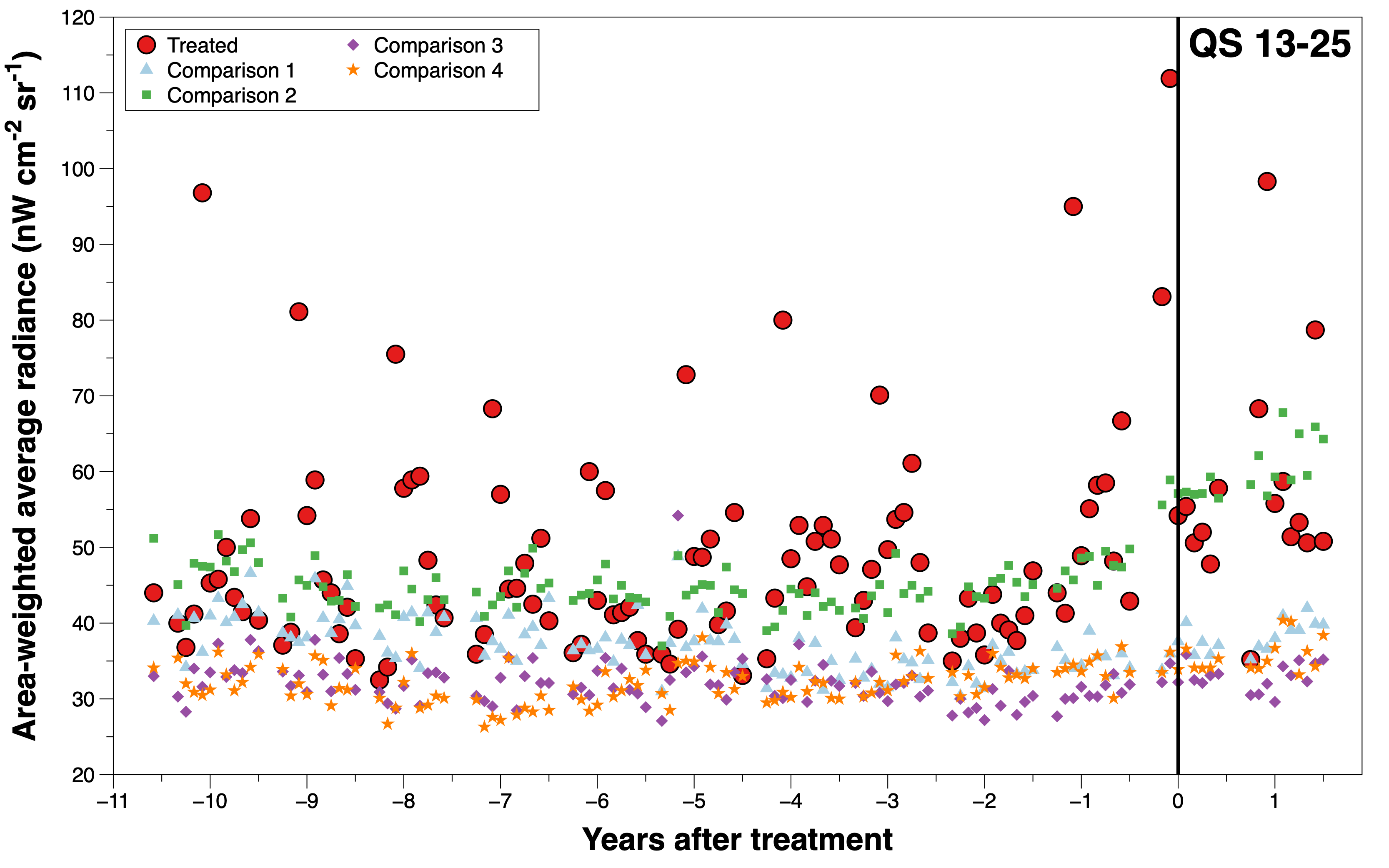}%
        }
        \caption{Upward radiance time series for Phoenix, Arizona, residential neighborhoods receiving CPT (`Treated') and control neighborhoods not receiving CPT (`Comparison' 1, 2, 3 and 4). Details are the same as in Figure~\ref{radiance-plots-1}.}
        \label{radiance-plots-3}
\end{sidewaysfigure*}

%
%
\begin{sidewaysfigure*}[t!]
\centering
        \subfloat[]{%
            \includegraphics[width=.48\linewidth]{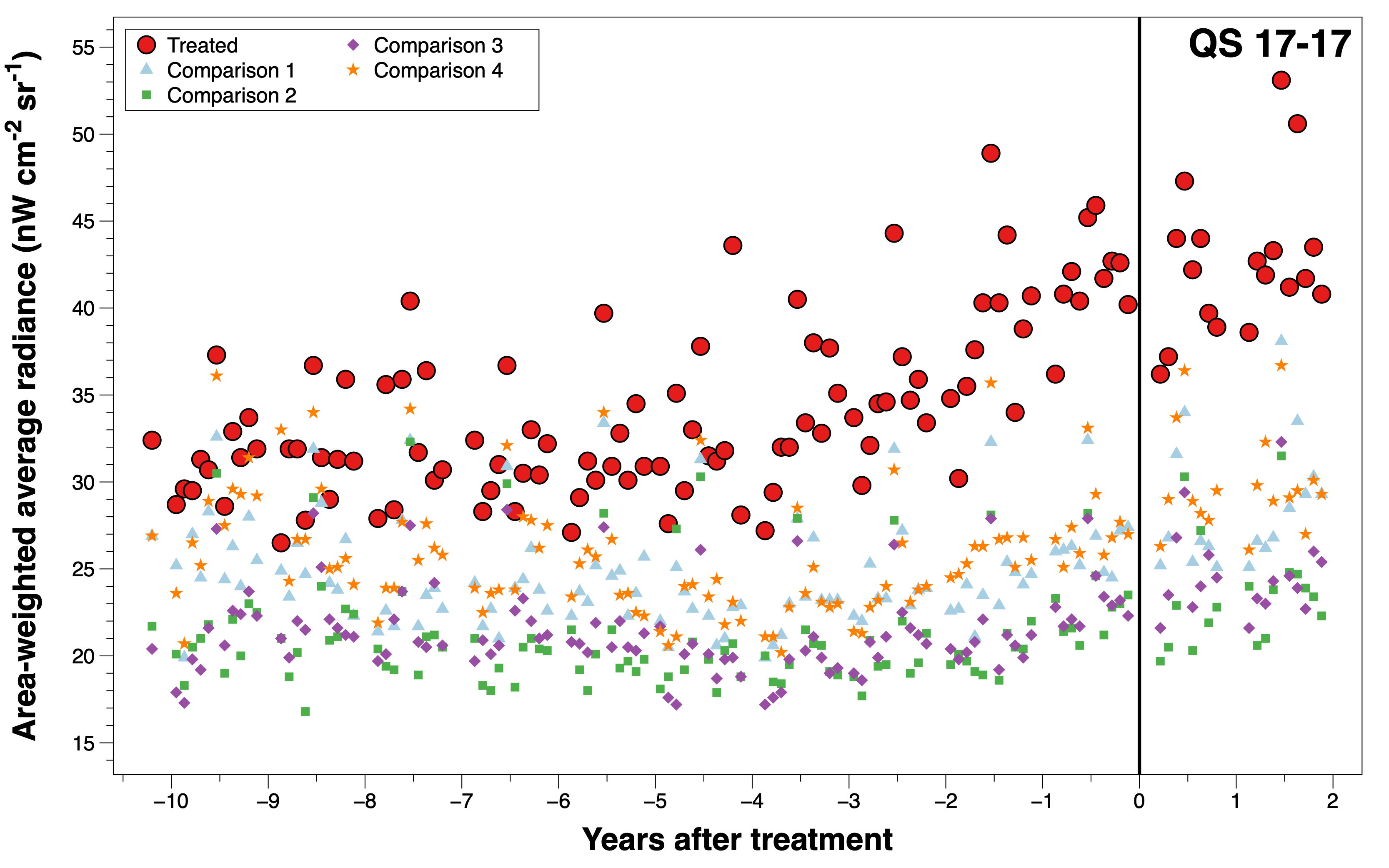}%
        }\hfill
        \subfloat[]{%
            \includegraphics[width=.48\linewidth]{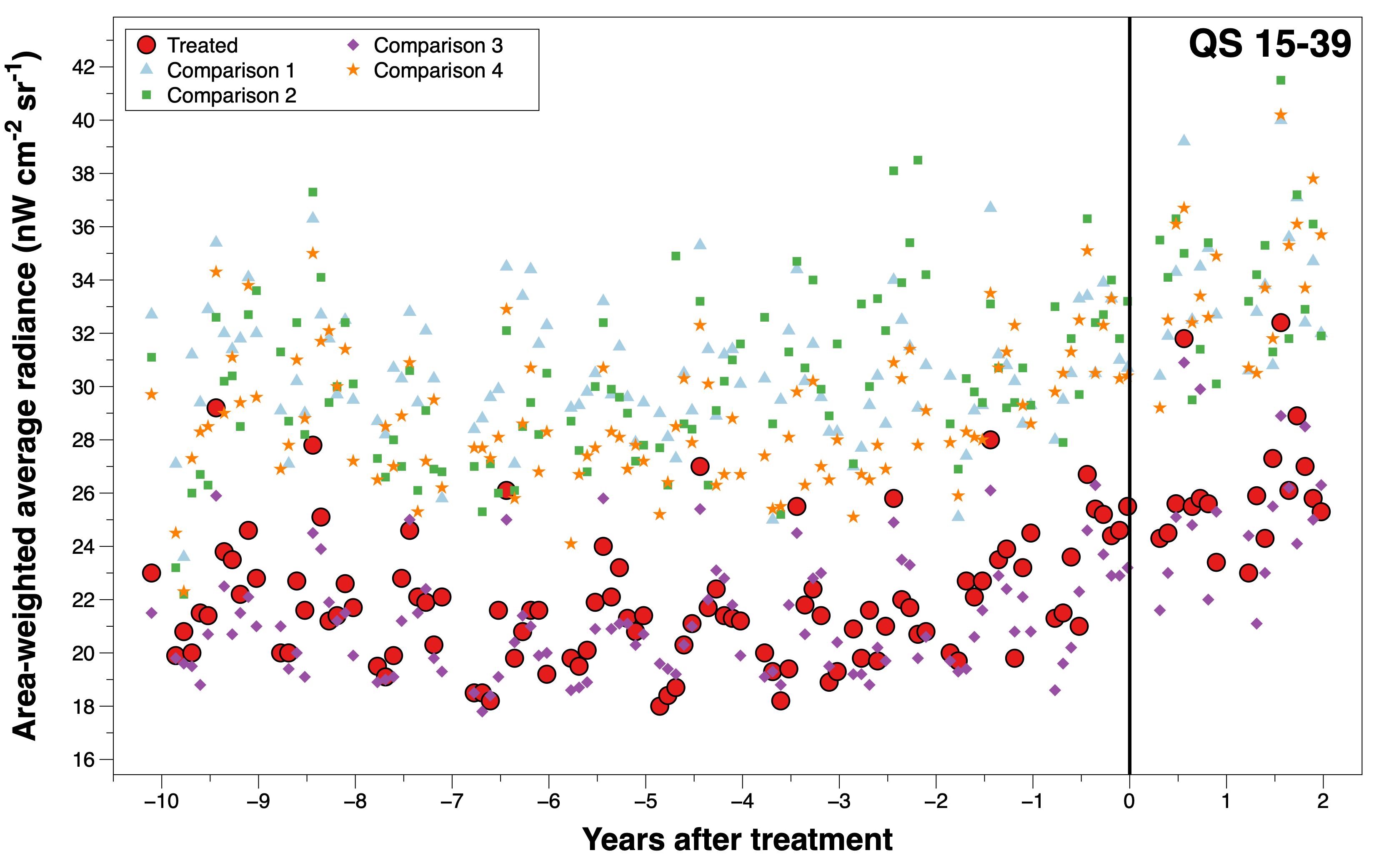}%
        }\\
        \subfloat[]{%
            \includegraphics[width=.48\linewidth]{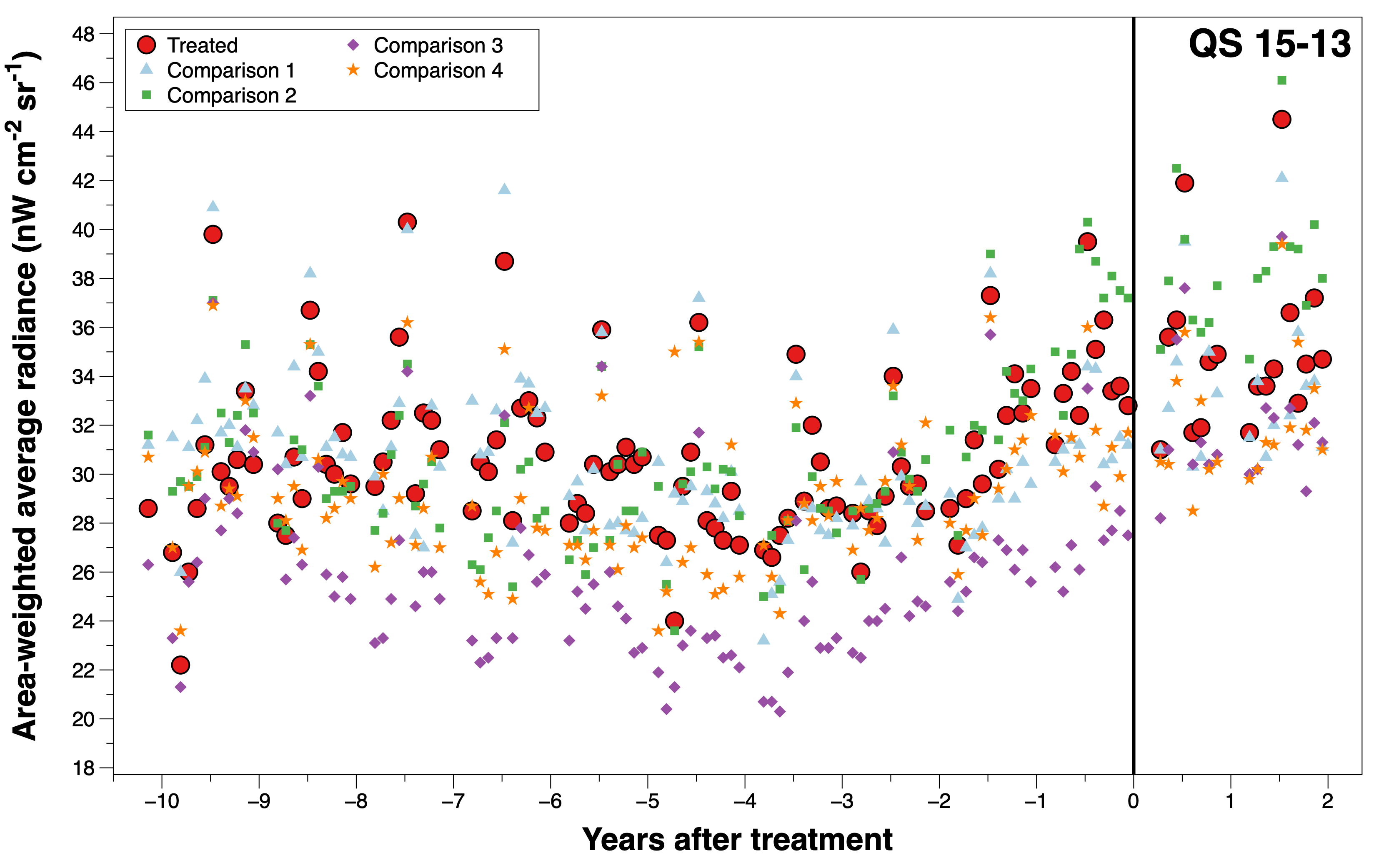}%
        }\hfill
        \subfloat[]{%
            \includegraphics[width=.48\linewidth]{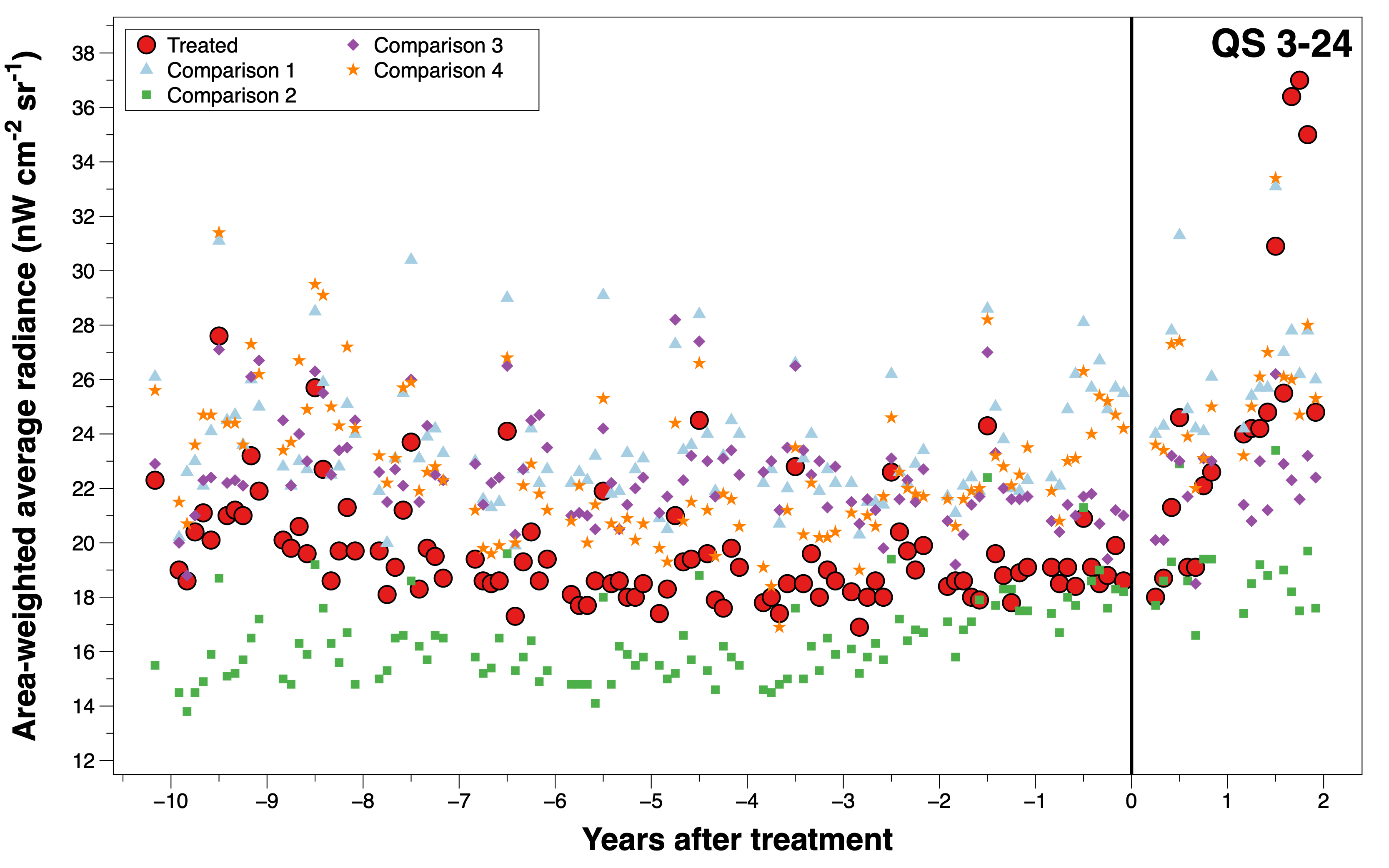}%
        }
        \caption{Upward radiance time series for Phoenix, Arizona, residential neighborhoods receiving CPT (`Treated') and control neighborhoods not receiving CPT (`Comparison' 1, 2, 3 and 4). Details are the same as in Figure~\ref{radiance-plots-1}.}        
        \label{radiance-plots-4}
\end{sidewaysfigure*}

%
%
\begin{sidewaysfigure*}[t!]
\centering
        \subfloat[]{%
            \includegraphics[width=.48\linewidth]{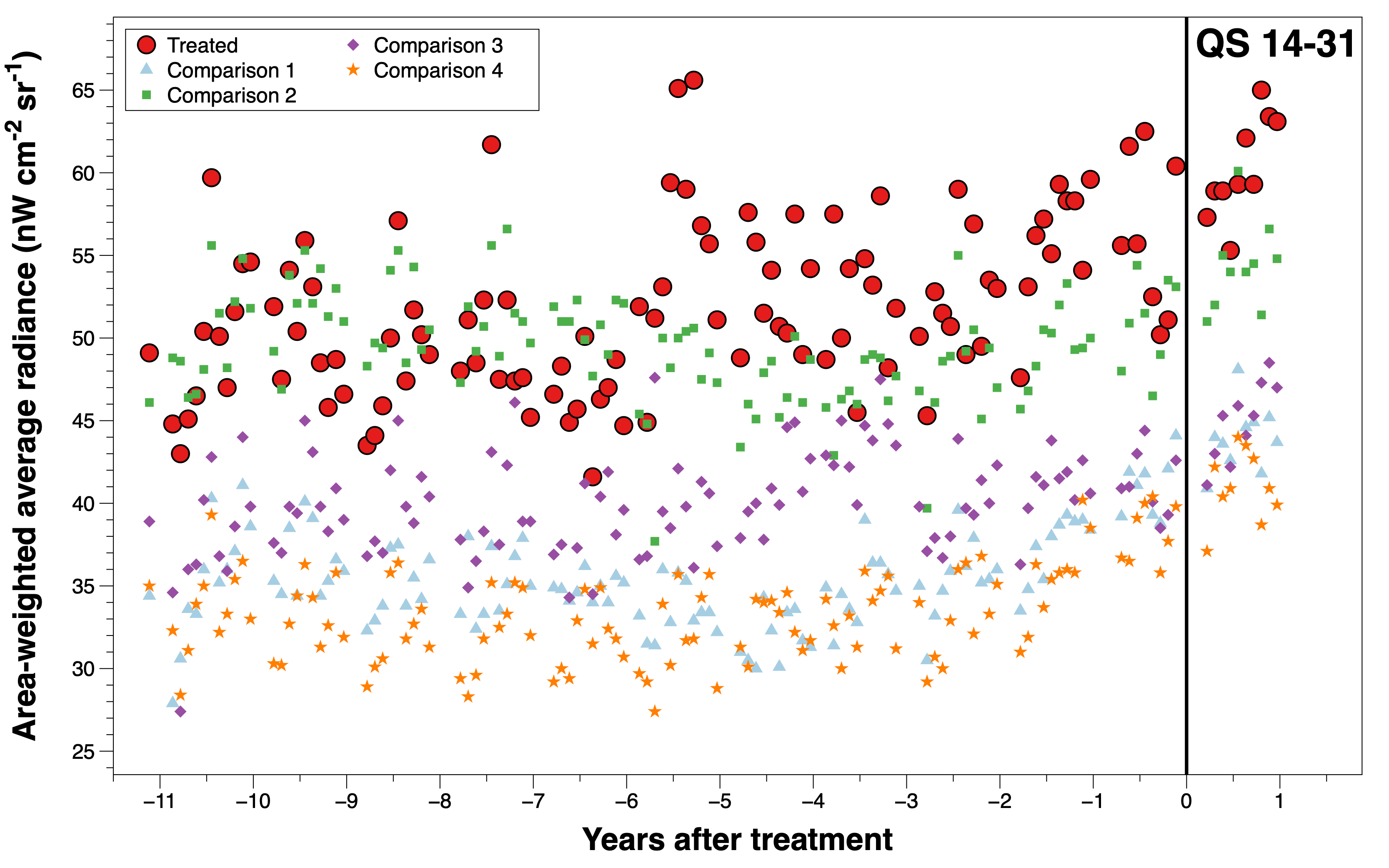}%
        }\hfill
        \subfloat[]{%
            \includegraphics[width=.48\linewidth]{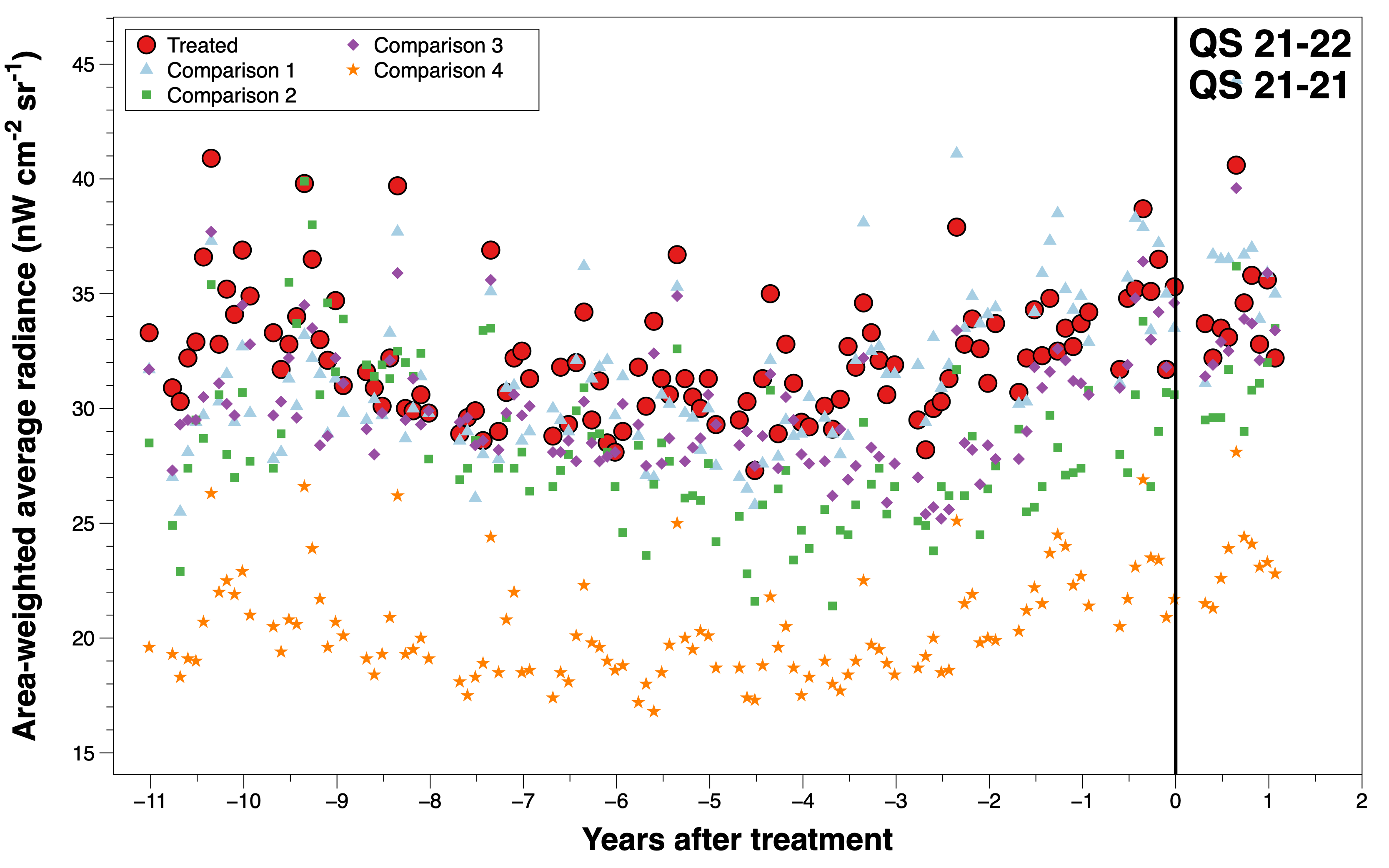}%
        }\\
        \subfloat[]{%
            \includegraphics[width=.48\linewidth]{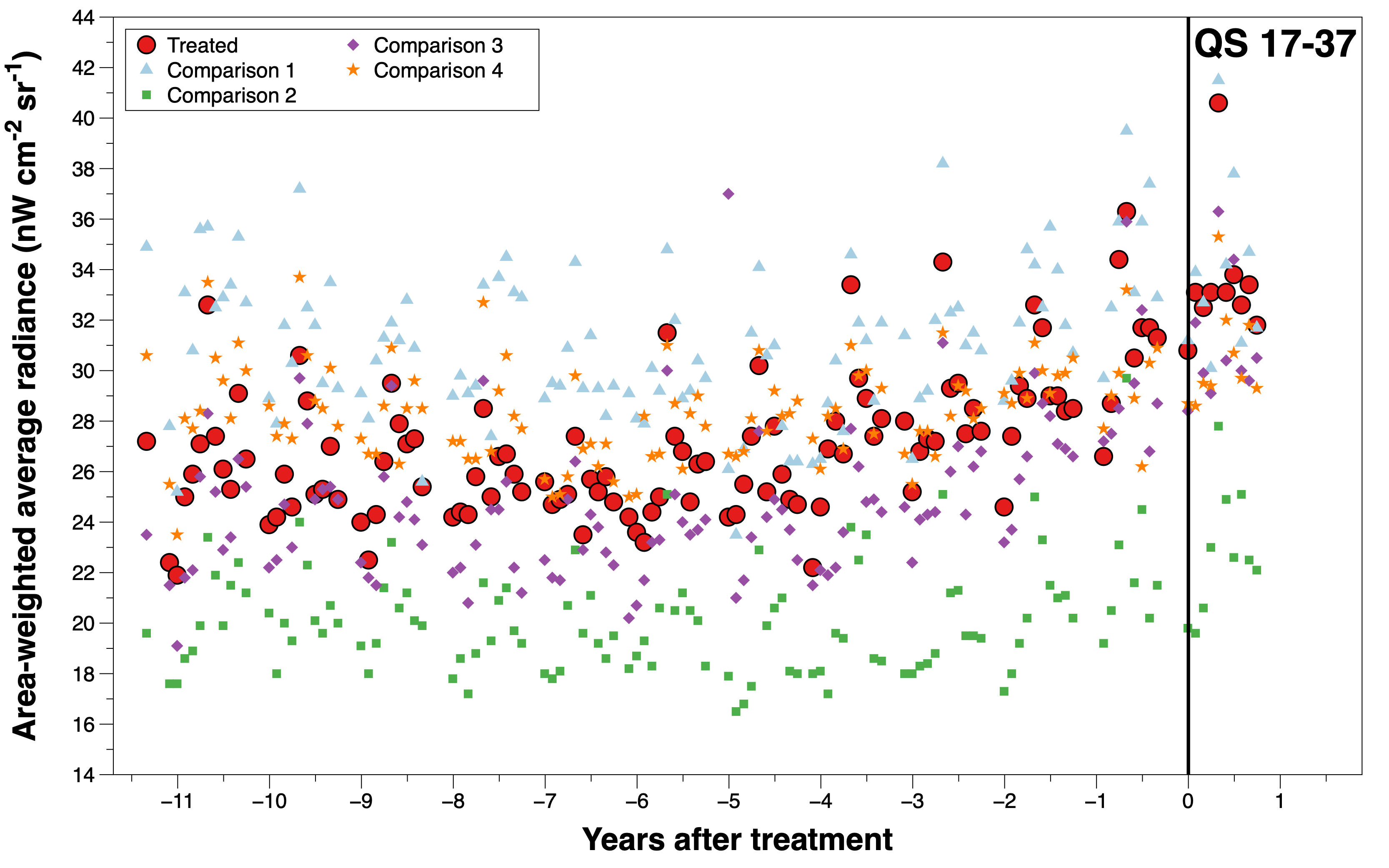}%
        }\hfill
        \subfloat[]{%
            \includegraphics[width=.48\linewidth]{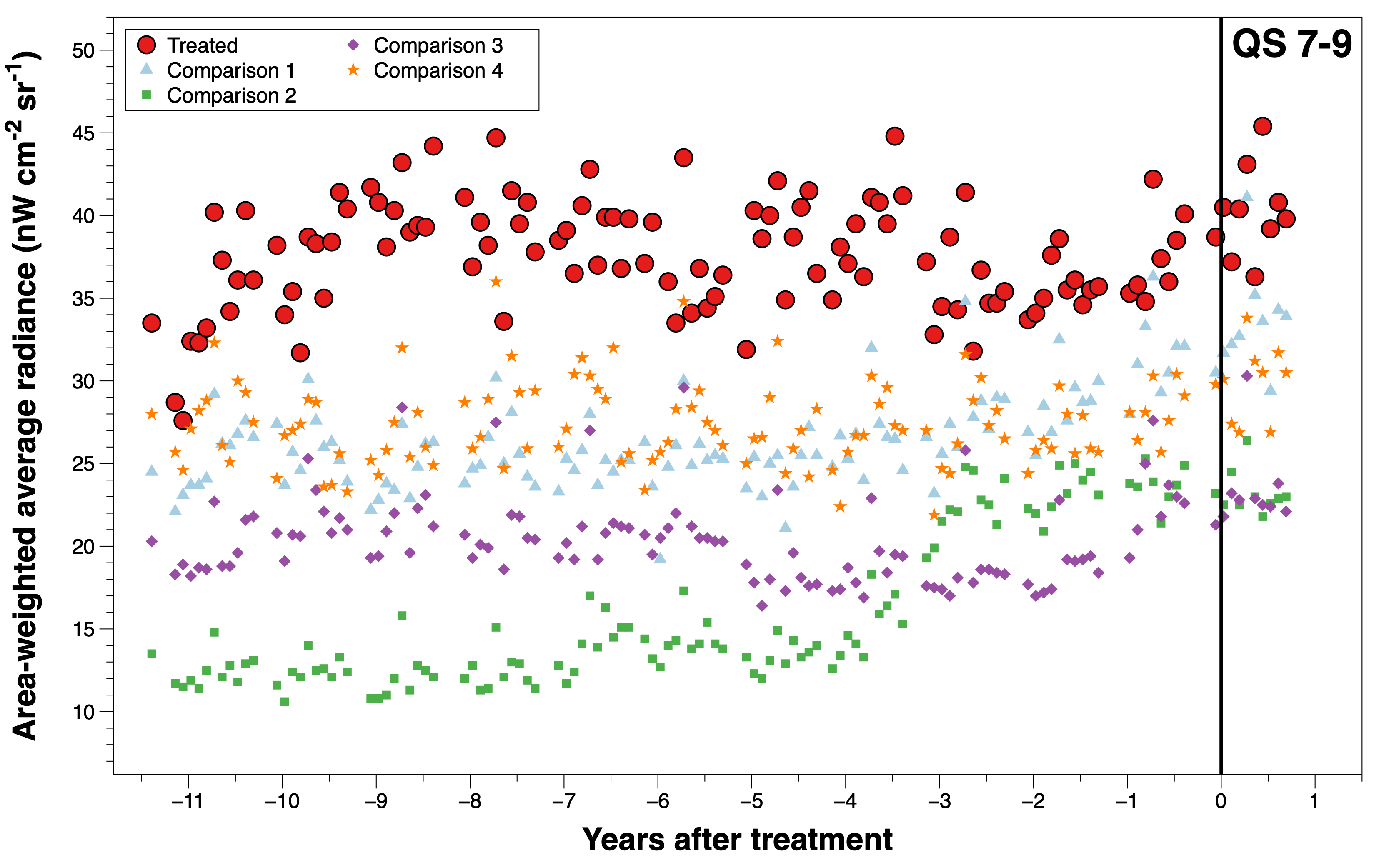}%
        }
        \caption{Upward radiance time series for Phoenix, Arizona, residential neighborhoods receiving CPT (`Treated') and control neighborhoods not receiving CPT (`Comparison' 1, 2, 3 and 4). Details are the same as in Figure~\ref{radiance-plots-1}.}        
        \label{radiance-plots-5}
\end{sidewaysfigure*}

%
%
\begin{figure}
\centering
\includegraphics[width=0.8\textwidth]{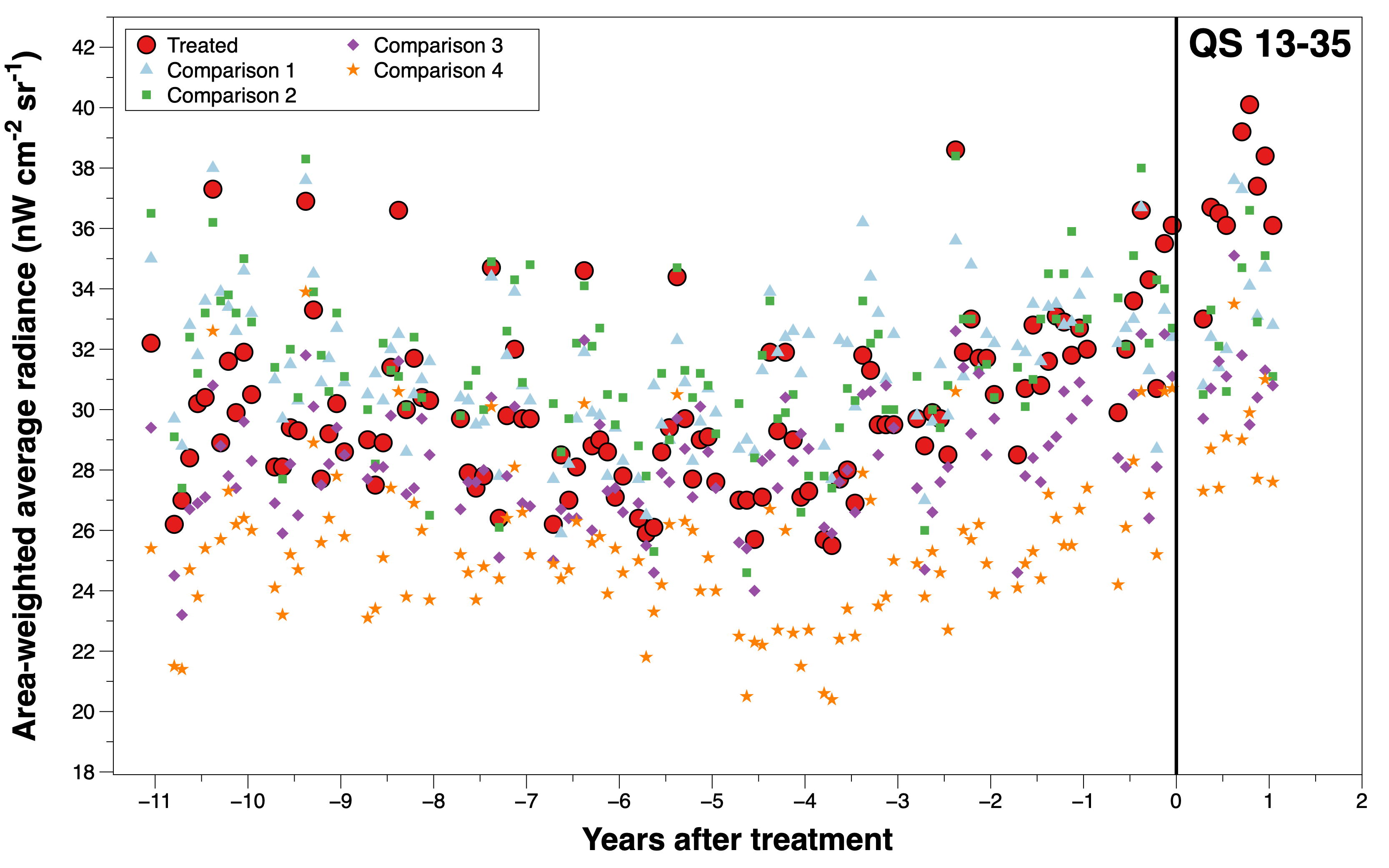}
\caption{Upward radiance time series for the Phoenix, Arizona, residential neighborhood labeled by the City of Phoenix `QS 13-35' receiving CPT (`Treated') and control neighborhoods not receiving CPT (`Comparison' 1, 2, 3 and 4). Details are the same as in Figure~\ref{radiance-plots-1}.}        
        \label{radiance-plots-6}
\end{figure}

\subsection{General description of the data}
\label{subsec:description}
The results of the process described in the previous section are presented in Figures~\ref{radiance-plots-1} through~\ref{radiance-plots-6}. The order of the panels in each figure, and in the series of figures overall, follows the temporal order in which CPT was applied in the treated neighborhoods from August 2020 (QS 33-18) to May 2023 (QS 13-35). There is considerable scatter in the data after removing secular trends evident in the long-baseline radiance time series. This is consistent with reports in the literature of standard deviations of as much as 15--20\% of the mean in long DNB time series.~\cite{Coesfeld2018} Month-to-month radiance changes in pairs of contiguous pixels are strongly correlated over spatial scales of a few kilometers, comparable to the sizes of the DNB pixel clusters we used to extract our time series. Changing the size of the sample region pixel clusters by up to a factor of about two in either the treated or comparison areas did not have any obvious effect on the statistics of the resulting time series.

There are a few excursions in the time series that are readily explainable. For example, in essentially every residential neighborhood measured we saw an isolated spike in radiance during the month of December. This is almost certainly due to the American tradition of deploying installations of decorative outdoor lighting each year around the time of the Christmas holiday. This is consistent with previous observations reported in the literature.~\cite{Roman2015,Ramirez2023} 

The reasons for other trends in our time series are not as clear. Many Phoenix neighborhoods show increases over the roughly 12-year time baseline of the series. Although a few neighborhoods' radiances were stable or slightly lower during the study period, some grew at rates faster than the recent rate of population change during the same time period (about +1\% year$^{-1}$).

It is important to note that the observed trends do not completely characterize the likely change that took place between 2012 and 2023. One significant influence is the ongoing modernization of lighting installations across the U.S., which involves replacing earlier technologies with mostly white LED lighting products. The relative insensitivity of the DNB to the blue light produced by these sources is known to result in undercounting radiances in areas where white LED dominates.~\cite{Cao2014} Short of conducting detailed inventories of outdoor lighting in the neighborhoods in this study using field data collection techniques, it is not possible to know the extent to which this effect influences the radiance time series.

\subsection{Predicted radiance changes due to CPT material application}
\label{subsec:predictions}
By developing the model described in Appendix A, we computed the expected change in radiance as a result of applying CPT materials in the treated neighborhoods using three pieces of information: the surface area subjected to the treatment and the reflectivity of the roadway surface before and after application of CPT. To determine the treated surface area in a given neighborhood we used information from the City of Phoenix on the length of the treated roadways and the typical width of residential streets in Phoenix (30 feet, or about 9 meters from curb to curb).~\cite{COPSPDG}

Schneider et al.~\cite{Schneider2023} report spectroradiometer measurements of treated surfaces in several Phoenix neighborhoods among the initial group receiving CPT between August and October 2020. The data were collected from November 2020 to May 2021 and are referred to the solar spectrum, reported in terms of percent reflectivity of that SPD. At optical wavelengths, they found that the average solar reflectivity of all eight neighborhoods diminished steadily by a total of about 10\% over a period of six months, at which point the treated surfaces showed only marginally higher reflectivity than untreated asphalt concrete. Table~\ref{radiance-change-predictions} shows the resulting predictions for the radiance change attributable to CPT material application based on the Schneider et al.~data.
\begin{table}
\centering
\begin{tabular}{ lccccc }
\hline
\textbf{} 	     & \textbf{Treated} 	 & \textbf{} 			    	     & \textbf{Area of Sample} & \textbf{} & \textbf{Expected} \\
\textbf{Code} & \textbf{area (m$^2$)} & \textbf{$N_{\textrm{DNB pix}}$} & \textbf{Region (m$^2$)} & \textbf{$\epsilon$} & \textbf{${\Delta}L/L(1)$} \\ 
\hline
QS 33-18		&	51652	&	6	&	1074150	&	0.05	&	0.02	\\
QS 59-23		&	103011	&	12	&	2148300	&	0.05	&	0.02	\\
QS 29-37		&	81820	&	6	&	1074150	&	0.08	&	0.04	\\
QS 11-22		&	77699	&	9	&	1611225	&	0.05	&	0.02	\\
QS 18-16		&	38555	&	2	&	358050	&	0.11	&	0.05	\\
QS 26-28		&	71960	&	6	&	1074150	&	0.07	&	0.03	\\
QS 2-19		&	59599	&	9	&	1611225	&	0.04	&	0.02	\\
QS 4-25		&	52388	&	6	&	1074150	&	0.05	&	0.02	\\
QS 40-17		&	44883	&	2	&	358050	&	0.13	&	0.06	\\
QS 37-37		&	85499	&	6	&	1074150	&	0.08	&	0.04	\\
QS 38-31		&	56362	&	6	&	1074150	&	0.05	&	0.03	\\
QS 13-25		&	29432	&	4	&	716100	&	0.04	&	0.02	\\
QS 17-17		&	83291	&	4	&	716100	&	0.12	&	0.06	\\
QS 15-39		&	53860	&	6	&	1074150	&	0.05	&	0.03	\\
QS 15-13		&	59010	&	9	&	1611225	&	0.04	&	0.02	\\
QS 3-24		&	66221	&	9	&	1611225	&	0.04	&	0.02	\\
QS 14-31		&	44147	&	6	&	1074150	&	0.04	&	0.02	\\
QS 21-21/22	&	111840	&	10	&	1790250	&	0.06	&	0.03	\\
QS 17-37		&	56067	&	6	&	1074150	&	0.05	&	0.03	\\
QS 7-9		&	71666	&	6	&	1074150	&	0.07	&	0.03	\\
QS 13-35		&	74756	&	6	&	1074150	&	0.07	&	0.03	\\
\end{tabular}
\caption{Summary of predictions for the radiance change attributable to CPT material application in the treated Phoenix neighborhoods based on the model described in Appendix A. The treated area is calculated from the published length of treated streets in selected neighborhoods multiplied by the average width of residential streets in Phoenix. $N_{\textrm{DNB pix}}$ is the number of contiguous DNB pixels in the sample region for each neighborhood. $\epsilon$ is the treated fraction of the sample region relative to its total area, and ${\Delta}L/L(1)$ is the resulting expectation for the relative change in radiance from the sample region that results from CPT application using the mean value of the measured reflectivity difference for treated versus untreated surfaces reported in~\cite{Schneider2023}.}
\label{radiance-change-predictions}
\end{table}

\subsection{Hypothesis testing}
\label{subsec:hypothesis-testing}
In order to test the hypothesis that applying CPT materials to asphalt concrete pavements increase the nighttime upward radiances of treated neighborhoods, it is useful to consider which changes in the top-of-the-atmosphere radiance can be detected with the DNB. The LOESS fits described in Section~\ref{sec:method} represent the measured radiances as a function of time, $L_{\textrm{DNB}}$($t$). They have uncertainties $\sigma_{\textrm{DNB}}$. The difference in the measured radiance just before and after CPT application is
\begin{equation}
{\Delta}L_{\textrm{DNB}} = L_{\textrm{DNB}_{\textrm{after}}}-L_{\textrm{DNB}_{\textrm{before}}},
\end{equation}
and the uncertainty on this quantity is
\begin{equation}
\sigma_{{\Delta}L_{\textrm{DNB}}} = \sqrt{{\sigma^2_{\textrm{DNB}_\textrm{after}}}+{\sigma^2_{\textrm{DNB}_\textrm{before}}}}.
\end{equation}
If the uncertainty of radiance measurements before and after CPT application is about the same, then $\sigma_{\textrm{DNB}_\textrm{after}}\approx\sigma_{\textrm{DNB}_\textrm{before}}\equiv\sigma_{\textrm{DNB}}$ such that $\sigma_{{\Delta}L_{\textrm{DNB}}}=\sqrt{2}\sigma_{\textrm{DNB}}$.

We want a measure of uncertainty that defines an interval about the measurement result within which the value of the measurement ${\Delta}L_{\textrm{DNB}}$ can be confidently asserted to lie. The combined standard uncertainty, $u$, is used to express the uncertainty of many kinds of measurement results. In this case, $u=\sigma_{{\Delta}L_{\textrm{DNB}}}$. The measure of uncertainty needed, ``expanded uncertainty'' ($U$), is found by multiplying $u$ by a ``coverage factor'', $k$, such that $U=ku$. The unknown true value of ${\Delta}L_{\textrm{DNB}}$ is confidently assumed to be both $\geq {\Delta}L_{\textrm{DNB}}-U$ and $\leq {\Delta}L_{\textrm{DNB}}+U$. $k$ is the critical $t_{1-{\alpha/2},\nu}$ value of the Student's $t$ distribution for a significance level $\alpha$ and $\nu$ degrees of freedom. When $\nu$ is large, and assuming that the data follow an underlying normal distribution, $k=2$ approximates 95\% coverage; that is, it represents the 95\% confidence interval (CI). 

The null hypothesis is ${\Delta}L_{\textrm{DNB}}={\Delta}L_{\textrm{0}}$, for which ${\Delta}L_{\textrm{0}}=0$ is only one possibility among many. Either of the following scenarios may be true:
\begin{enumerate}
\item If $|{\Delta}L_{\textrm{DNB}}-{\Delta}L_{\textrm{0}}| \geq k\sigma_{{\Delta}L_{\textrm{DNB}}}=k\sqrt{2}\sigma_{\textrm{DNB}}$, then the null hypothesis is rejected for the CI associated with $k$ because the probability of measuring ${\Delta}L_{\textrm{DNB}}$ given that the ``true'' value is ${\Delta}L_{\textrm{0}}$ is small ($<1-\textrm{CI} = 0.05$), or else
\item If $|{\Delta}L_{\textrm{DNB}}-{\Delta}L_{\textrm{0}}| < k\sigma_{{\Delta}L_{\textrm{DNB}}}=k\sqrt{2}\sigma_{\textrm{DNB}}$, then the null hypothesis cannot be excluded at the stated CI because the probability of measuring ${\Delta}L_{\textrm{DNB}}$ given that the true value is ${\Delta}L_{\textrm{0}}$ is $>1-\textrm{CI} = 0.05$.
\end{enumerate}
In the second case, there may be other values of ${\Delta}L_{\textrm{0}}$ that also make the inequality true, but all values of $k$ cannot make it true at once. One can only confidently assert that ${\Delta}L_{\textrm{DNB}}$ differs from ${\Delta}L_{\textrm{0}}$ by an amount smaller than a critical value that represents the criterion for discarding the null hypothesis.

This is not quite the case because $\sigma_{{\Delta}L_{\textrm{DNB}}}$ ``blurs'' perceiving changes that could contradict the null hypothesis. An example is ${\Delta}L_{\textrm{0}}=0$, which is contradicted only if one measures $|{\Delta}L_{\textrm{DNB}}| \geq k\sigma_{{\Delta}L_{\textrm{DNB}}}$. If, on the other hand, $|{\Delta}L_{\textrm{DNB}}| < k\sigma_{{\Delta}L_{\textrm{DNB}}}$, the usual conclusion is that the change is not statistically significant. When $\sigma_{{\Delta}L_{\textrm{DNB}}}$ is large relative to $|{\Delta}L_{\textrm{DNB}}|$, it is not possible to confidently claim significance. Yet a change in radiance deemed insignificant could still yield a non-trivial contribution to skyglow over an area. 

We applied this method to the radiance time series for the neighborhoods receiving CPT. The hypothesis tested in each case was the radiance change prediction described in Section~\ref{subsec:predictions}. The results of these tests are given below in Table~\ref{hypothesis-tests}. 
\begin{table}
\centering
\begin{tabular}{ lcccccc }
\hline
\textbf{} & \textbf{} 	 & \textbf{}  & \textbf{} & \textbf{} & \textbf{} & \textbf{Hypothesis} \\
\textbf{Code} & \textbf{$L_{\textrm{before}}$} & \textbf{$L_{\textrm{after}}$} & \textbf{${\Delta}L_{\textrm{DNB}}$} & \textbf{${\Delta}L_{\textrm{0}}$} & \textbf{$k\sigma_{{\Delta}L_{\textrm{DNB}}}$} & \textbf{Excluded?} \\ 
\hline
QS 33-18		&	21.6		&	20.7	&	-0.9	&	0.5	&	2.1	&	Yes	\\
QS 59-23		&	10.9		&	10.9	&	0.0	&	0.3	&	1.5	&	Yes	\\
QS 29-37		&	15.1		&	16.1	&	1.0	&	0.4	&	1.6	&	Yes	\\
QS 11-22		&	25.6		&	27.9	&	2.3	&	0.6	&	2.4	&	Yes	\\
QS 18-16		&	28.2		&	31.1	&	2.9	&	1.5	&	2.6	&	Yes	\\
QS 26-28		&	36.2		&	38.8	&	2.6	&	1.2	&	2.0	&	Yes	\\
QS 2-19		&	15.1		&	15.7	&	0.6	&	0.3	&	0.9	&	Yes	\\
QS 4-25		&	17.0		&	17.6	&	0.7	&	0.4	&	1.5	&	Yes	\\
QS 40-17		&	28.1		&	27.5	&	-0.6	&	1.8	&	2.6	&	Yes	\\
QS 37-37		&	43.1		&	33.5	&	-9.6	&	1.7	&	2.9	&	Yes	\\
QS 38-31		&	22.7		&	20.4	&	-2.3	&	0.6	&	1.7	&	Yes	\\
QS 13-25		&	111.9	&	54.2	&	-57.7	&	2.3	&	11.6	&	Yes	\\
QS 17-17		&	40.2		&	36.2	&	-4.0	&	2.3	&	3.9	&	Yes	\\
QS 15-39		&	25.5		&	24.3	&	-1.2	&	0.6	&	1.9	&	Yes	\\
QS 15-13		&	32.8		&	31.0	&	-1.8	&	0.6	&	2.8	&	Yes	\\
QS 3-24		&	18.6		&	18.0	&	-0.6	&	0.4	&	1.5	&	Yes	\\
QS 14-31		&	60.4		&	57.3	&	-3.1	&	1.2	&	4.5	&	Yes	\\
QS 21-21/22	&	35.3		&	33.7	&	-1.6	&	1.1	&	2.1	&	Yes	\\
QS 17-37		&	33.1		&	32.5	&	-0.6	&	0.9	&	1.9	&	Yes	\\
QS 7-9		&	38.7		&	40.5	&	1.8	&	1.3	&	2.9	&	Yes	\\
QS 13-35		&	36.1		&	33.0	&	-3.1	&	1.3	&	2.1	&	Yes	\\
\end{tabular}
\caption{Results of testing the hypothesis that observed DNB radiance changes in Phoenix neighborhoods receiving CPT are consistent with the predictions of the modeling described in Section~\ref{subsec:predictions} with 95\% confidence. Columns 2-5 show DNB radiance values in units of nW cm$^{-2}$ sr$^{-1}$.}
\label{hypothesis-tests}
\end{table}
We can exclude the hypothesis for all neighborhoods in the sample at the 95\% confidence level except for QS 4-25. Two neighborhoods, QS 18-16 and QS 26-28, show apparent radiance increases before and after CPT application. However, in neither case does the change equal or exceed the predicted increase when the observed scatter in LOESS fit residuals is taken into account. 

We also applied the same procedure to evaluate the radiance time series from the comparison areas not receiving CPT using the ``no change'' (${\Delta}L_{\textrm{0}}=0$) test criterion during the same time interval in which the corresponding treated area received CPT. In 78 of 84 comparison areas, the null hypothesis could not be excluded. In the six instances where it was, we found it was equally likely that a radiance \emph{decrease} exceeding the DNB time series noise occurred compared to a radiance increase.

We briefly comment on the result for one of the treated neighborhoods in particular, QS 13-25. This area is not representative of most Phoenix residential neighborhoods as it is situated immediately adjacent to the Arizona State Fairgrounds, which are partially enclosed by the DNB pixels used to generate the radiance time series used in this work. As seen in panel (d) of Figure~\ref{radiance-plots-3}, large radiance excursions occur semi-regularly due to public events held at the fairgrounds involving temporary uses of high-intensity outdoor lighting. This explains the very large value of ${\Delta}L_{\textrm{DNB}}$ in column 4 of Table~\ref{hypothesis-tests}, in which the time interval between radiance measurements included high light emissions during the annual Arizona State Fair. We found no comparable large deviations in the time series among the comparison areas for this neighborhood.

\subsection{Main finding}
\label{subsec:finding}
For the chosen 95\% confidence interval, we find that none of the Phoenix neighborhoods treated with CPT materials showed any increase in radiance consistent with the predictions of our models. Based on the known properties of the DNB, any increase in radiance attributable to CPT deployment consistent with the simple hypothesis articulated in Section~\ref{sec1:intro} appears to be below the detection threshold using DNB time series as the data source.

\subsection{Assumptions}
\label{subsec:assumptions}
A number of assumptions underlie our understanding of the main finding. The first is which kind of light source is most prevalent in the residential neighborhoods of Phoenix. Anecdotally, street lighting dominates light emissions in those neighborhoods during the overnight hours. In 2017-2020, the City replaced its stock of 92,500, mostly partially shielded high-pressure sodium (HPS) roadway luminaires with fully shielded, 2700 K white LED roadway luminaires from a single manufacturer. The LED luminaires have nearly the same luminous flux as the HPS luminaires they replaced. As a matter of policy the City of Phoenix places roadway luminaires with a pole spacing of 60-76 m along ``local'' streets, which characterize the residential neighborhoods of the city.~\cite{COPSLLG} Other lighting in residential areas consists mainly of area and building egress applications, and much of this lighting is either partially shielded or unshielded. A distinct exception to this pattern is decorative holiday lighting deployed extensively between mid-November and early January of each year. This effect is readily noticeable in the radiance time series shown in Figures~\ref{radiance-plots-1} to~\ref{radiance-plots-6}. 

Because the new roadway luminaires are fully shielded, we assume that most of the upward-directed light in these neighborhoods results from Lambertian scattering from road surfaces consistent with previous field and laboratory measurements, e.g.,~\cite{Rossi2018}. Because the light distribution from the roadway luminaires includes a certain amount of backlight, we expect some upward light scatter from sidewalks adjacent to roadways. Where sidewalks exist, the City of Phoenix Streetlighting Layout Guidelines call for a 1-foot (30 cm) setback from the curb; otherwise, the setback from the curb is three feet (approximately 1 m).~\cite{COPSLLG} Scattering from sidewalks and other surfaces, and therefore its contribution to upward radiance, is assumed to be essentially invariant night to night. There is essentially never any snow or ice cover on roadways that would increase their albedos. Furthermore, given the desert climate of Phoenix, there is little in the way of trees or landscaping that shades streets from illumination by roadway lighting or absorbs reflected light that would otherwise be directed upward into the night sky. There is some presence of cars parked on the streets that have their own reflective properties, but their numbers and locations from night to night are random. We assume that they do not account for a significant fraction of upward radiance emanating from residential neighborhoods. 

\subsection{Interpretation of the finding}
\label{subsec:interp}
There are a number of possible explanations for why no statistically significant change points are observed in the radiance time series. 

First, CPT is typically applied to worn asphalt. Asphalt concrete is spectrally red, and the spectral slope in the blue gets very steep with age. This material becomes more reflective at all wavelengths as binders wear away, exposing higher-albedo matrix materials; see, e.g., Figure 1 in~\cite{Herold2004}. Compared to untreated asphalt concrete, especially when the untreated material is worn from environmental exposure, the CPT material does not substantially increase the albedo of surfaces at optical wavelengths. 

Second, the spectral properties of CPT materials are similar to those of asphalt concrete, accounting for in situ aging effects. The City of Phoenix CPT material is also slightly red when newly applied,~\cite{COPE2023} and both treated surfaces and untreated asphalt concrete appear to ‘age’ in a spectrally neutral way.~\cite{Schneider2023} Asphalt itself is one of the main components of the CPT materials, and the other components responsible for its reflective characteristics are more efficient at infrared wavelengths. 

And third, the DNB is not especially sensitive to these very red materials, so changes in their reflectivity at long wavelengths are not detectable in the radiance time series. The DNB spectral sensitivity has changed continuously since the original deployment of VIIRS. Its relative red response has degraded over time, while its relative blue response has improved.~\cite{Lee2014} This complicates the interpretation of radiance time series for light sources that have steep spectral slopes at optical wavelengths.

\begin{figure}[ht]
\centering
\includegraphics[width=.95\linewidth]{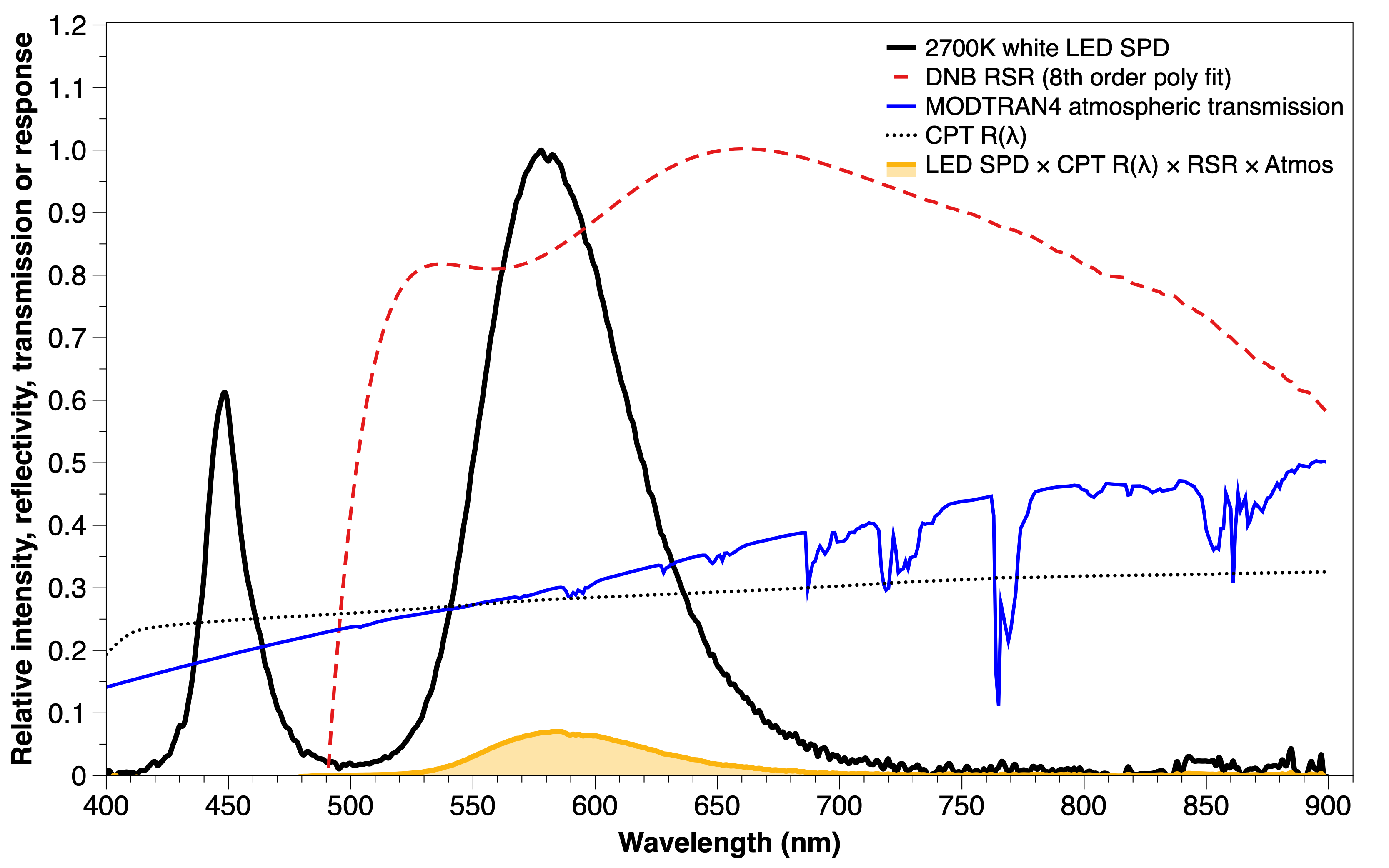}
\caption{A representation of the effective spectral content of the signal recorded by the DNB in the direction of Phoenix neighborhood receiving CPT. The spectral power distribution of a generic 2760 K (nominal) white LED lamp (solid black line, from~\cite{LSPDD}) is convolved with (1) the spectral reflectivity of freshly applied Phoenix CPT material measured in situ (dotted black line, from~\cite{COPE2023}); (2) a polynomial fit to the predicted relative spectral response (RSR) of the DNB seven years after launch (red dashed line, from~\cite{Lei2013}); and (3) a MODTRAN4 atmospheric transmission model for a summertime mid-latitude hazy atmosphere (solid blue line, from~\cite{MODTRAN4}). The result (solid yellow line with yellow shading below it) is what the DNB would detect if the only source of upward radiance in treated neighborhoods were the Lambertian reflection of street lighting from newly treated roadway surfaces.}\label{CPT-lamp-SPD-convolution}
\end{figure}
Figure~\ref{CPT-lamp-SPD-convolution} shows what we interpret the DNB to ``see'' when it looks at treated neighborhoods. To compute the effective spectral intensity of the upward-directed light, we assumed that the source was limited to streetlights shining on recently applied CPT materials. The source emits light with a SPD characteristic of blue-pump white LED, consisting of a ``blue spike'' near 450 nm and a broader peak centered around 580 nm resulting from phosphor conversion of blue light to other colors. This light is processed by several influences before detection. It is reflected from the CPT material, passes upward through the atmosphere, and is detected by a sensor whose response to light as a function of wavelength is known to change as it ages.~\cite{Lei2013} The SPD of the source is shown in Figure~\ref{CPT-lamp-SPD-convolution} along with the spectral distributions of the modifying influences and the result after convolving them with the source. That result is the expected SPD of light from Phoenix neighborhoods if that light consisted only of street lighting reflections from recently applied CPT materials with no contributions from other sources. In reality, there are unknown contributions to the total radiance signal from domestic lighting and reflections of street lighting from surfaces other than roadways. 

\subsection{Confounding influences}
\label{subsec:confounders}
There are a number of reasons why the radiance data we show here could mask actual changes to nighttime upward light emissions in Phoenix neighborhoods receiving CPT. The DNB pixels are relatively large, and in cities they tend to dilute the signal associated with any particular, spatially small lighting installation. The treated Phoenix neighborhoods all subtend no more than nine contiguous DNB pixels. However, given the relatively late overpass time of the NPP satellite and that upward light emissions during those hours are dominated by reflected street lighting, we find this scenario unlikely. 

Instrumentally scattered light also could mask the effect of radiance changes due particularly to CPT application.~Petržala and Kómar~\cite{Petrzala2024} recently showed that in situations where the aerosol optical depth in the direction of observation is high, scattered light could account for as much as 10-20\% of the measured radiance in a given DNB pixel. Additionally, flicker caused by outdoor lighting powered by alternating current (AC) may affect DNB radiances.~\cite{Elvidge2022} Signals from adjacent pixels are not completely uncorrelated, which suggests that at least some lighting AC phases are shared between them. This effect may modify the radiance ratios before and after CPT application as seen by the DNB.  

Another possibility is that prompt albedo increases after CPT is applied may be removed quickly by environmental exposure effects. In this way, any contribution to increased upward light emission immediately after CPT application may decline so quickly as to simply not influence radiances derived from monthly composite images made around the same time. The in-situ spectroradiometric measurements of the Phoenix CPT materials made by Schneider et al.~\cite{Schneider2023} mentioned previously indicated a steady decline in spectral reflectivity after application. The observed decline is within the uncertainty of the detection threshold of our radiance time series, so we cannot rule out this scenario.

Lastly, it may be the case that discontinuities in the radiance time series masked an albedo change because short-term increases took place during times when DNB radiances from monthly cloud-free composite images were not available. In the radiance time series for the Phoenix neighborhoods presented in this paper, there is often no data available for summer months, particularly June through July when nights are short in the northern hemisphere. This is the result of a conscious choice by the developers of RLT to exclude these months given concerns about inadequate subtraction of scattered light. By coincidence, some CPT were applied in months adjacent to these, leading to gaps in the time series that would otherwise tend to reveal changes. We cannot rule out the possibility in those instances that real radiance changes were not noticed due to data being unavailable.

\subsection{Testing the hypothesis by means of skyglow measurements}
\label{subsec:NSB}
One may well ask whether time-series measurements of skyglow might reveal a change difficult to perceive in the radiance time-series. To the extent that some directed upward after reflecting from surfaces treated with CPT is scattered during its flight through the atmosphere to become skyglow, it is reasonable to expect that the brightness of the night sky should change proportionately. In isolation, this effect is likely to be observed. And indeed, modern methods of measuring the brightness of the night sky can detect changes of the magnitude predicted by the model presented here in~\ref{sec:app1}.~\cite{Bara2019,Barentine2020,Robles2021,Fiorentin2022} But in the context of a sprawling urban area, it is difficult to detect small light emission changes as reflected in time series measurements of night-sky brightness from lighting interventions that have small spatial footprints. Given the nature of skyglow formation, the signal from such installations is diluted from the much stronger signal of the surrounding city. For this reason we assumed that any skyglow signal resulting from CPT application would be practically impossible to detect, and we therefore did not make any night-sky brightness measurements. While changes in skyglow of the expected magnitude resulting from CPT application might be possible to detect in the context of a geographically isolated village, for example, we know of no instances to date in which such an experiment could have been conducted.

%
%
\section{Summary and conclusions}
\label{sec:summary}

In this study we considered whether the application of reflective materials over paved roadways to reduce the heating of air as a means of helping cities adapt to climate change has any unintended side-effect of increasing the nighttime radiance of treated areas. In an effort to detect such an effect, we obtained radiance time series based on data from the Visible Infrared Imaging Radiometer Suite (VIIRS) aboard the Suomi National Polar-orbiting Partnership satellite. Radiances were measured in the wavelength range 500-900 nm with the VIIRS Day-Night Band (DNB). We used these time series to compare radiance changes in 22 residential neighborhoods in Phoenix, Arizona, U.S., receiving ``cool pavement'' treatments (CPT) with untreated control neighborhoods. 

We found no statistically significant change in the nighttime radiance of treated neighborhoods after CPT materials were applied. This conclusion takes into account that the minimum radiance change detectable as significant by our measurement system (at 95\% confidence) is $0.9 \leq k\sigma_{{\Delta}L_{\textrm{DNB}}} \leq 11.6$, or, expressed in percent increase above the initial radiance, $5.5\% \leq k\sigma_{{\Delta}L_{\textrm{DNB}}}/L_{\textrm{DNB}_{\textrm{before}}} \leq 14.1\%$. In each of the treated neighborhoods, the changes we measured were smaller than this. Given that the range of ${\pm}k\sigma_{{\Delta}L_{\textrm{DNB}}}$ can be relatively large, the existence of actual changes within this range cannot be excluded based on the data. While it is possible that there was in fact a change in the upward radiance due to the application of CPT consistent with our model within the range of ${\pm}k\sigma_{{\Delta}L_{\textrm{DNB}}}$, it is equally possible that there was no change. We therefore cannot disprove (reject) with 95\% confidence any null hypothesis that CPT produces a radiance change ${\Delta}L_0$ within the interval ${\Delta}L_{\textrm{DNB}}{\pm}k\sigma_{{\Delta}L_{\textrm{DNB}}}$, inclusive of the no-change hypothesis ${\Delta}L_0=0$.

From this we conclude that at least for the particular variety of CPT applied by Phoenix in 2021-2023 the expected radiance changes are confined within the interval ${\Delta}L_{\textrm{DNB}}{\pm}k\sigma_{{\Delta}L_{\textrm{DNB}}}$, or, in percent with respect to the pre-CPT values,
\begin{equation}
\frac{{\Delta}L_{\textrm{DNB}}}{L_{\textrm{DNB}_{\textrm{before}}}} \pm \frac{k\sigma_{{\Delta}L_{\textrm{DNB}}}}{L_{\textrm{DNB}_{\textrm{before}}}}.
\end{equation}
In this representation, values range from $-51.5\pm10.4$\% (QS 13-25) to $+10.3\pm9.2$\% (QS 18-16). As a consequence, we cannot discount the possibility that the future deployment of similar CPT materials in cities like Phoenix will modify urban skyglow by as much as the latter figure with respect to pre-treatment values. Given the limitations of the current measurement system (VIIRS-DNB radiances), we simply cannot make any stronger statements about the existence or not of actual changes that are even slightly smaller than this for the chosen confidence level. This should borne in mind when evaluating the results of other studies that depend on DNB time-series data. Nevertheless, it is reasonable to suggest that future applications of CPT in cities should be coupled with reductions in the illuminance yielded by lighting systems with the goal of holding the \emph{luminance} of surfaces constant through the transition. This will benefit all users of road networks and support public safety while ensuring that there is no significant change in the upward radiance of illuminated road surfaces that may in turn contribute to brightening the night sky.

Further investigation may reveal whether a real change in nighttime upward radiance from treated neighborhoods is masked by shortcomings of the DNB as a data source. Calibrated nighttime satellite imagery with higher spatial resolution, or measurements made from aircraft- or balloon-based platforms, may more clearly show the effects of CPT application. More temporally granular data such as the DNB nightly composites may resolve the question of whether an abrupt increase in roadway albedo from CPT application fades more quickly than the monthly cloud-free composites suggest. This may also improve the continuity of time series across application dates. 

A better-controlled and more definitive experiment could be conducted by applying CPT across an isolated geographic territory in which the spatial extent of outdoor lighting is comparatively small, and that territory is surrounded by a much larger region without lighting. This is akin to the situation in a rural village. But its spatial extent should be sufficient so as to subtend several DNB pixels, and all-sky photometry should be obtained from several vantage points in and beyond the area where lighting is installed. In this way more robust conclusions may be drawn from the available data, even if this scenario is less representative of real-world conditions.

\appendix
\section{Expected radiance changes due to CPT deployment}
\label{sec:app1}
For each DNB pixel, labeled $n$, covering a treated neighborhood there is a radiance representing the condition before CPT materials are applied (condition ``1''); that is, $L_{\textrm{DNB}_{1}} \equiv L_{n}(1)$. It consists of the sum of two contributions:
\begin{enumerate}
\item A background radiance $L_{n,bg}$ that is unrelated to artificial light sources in the pixel, in particular scattered light from adjacent DNB pixels and light scattered during transit through the atmosphere; and
\item Radiance from artificial light sources in the pixel $L_{n,p}$ that comes from both direct emission from source to detector ($L_{n,pd}$) and indirect emission due to reflections of light from surfaces within the pixel ($L_{n,pr}$). 
\end{enumerate}
For purely Lambertian reflections from surfaces within a pixel,
\begin{equation}
L_{n,pr} = \frac{\rho_n}{\pi}\frac{\Phi_n}{S_n},
\end{equation}
where $\rho_n$ is the average reflectance of those surfaces, $\Phi_n$ is total luminous flux reaching the surfaces, and $S_n$ is the spatial area of the pixel. $L_{n}(1)$ is therefore the sum of all the radiance contributions:
\begin{align}
L_{n}(1) &= L_{n,bg} + L_{n,pd} + L_{n,pr} \\
              &= L_{n,bg} + L_{n,pd} + \frac{\rho_n}{\pi}\frac{\Phi_n}{S_n}
\label{eqnA3}
\end{align}
In neighborhoods receiving CPT, the materials are applied to a fraction $\epsilon_n$ of the area of one pixel. The area receiving the treatment is taken as the width of the roadway multiplied by the length of the treated streets in a neighborhood, the result of which is divided by the spatial area of one DNB pixel. After CPT materials are applied (condition ``2''), the surface reflectance is defined as $\rho^{\prime}_n$. For constant luminous flux, presumed to be mostly from streetlights, reaching the roadway surface, the radiance originating from within the DNB pixel directed upward toward the top of the atmosphere is
\begin{equation}
L_{n}(2) = L_{n,bg} + L_{n,pd} + \left ( \frac{\Phi_n}{{\pi}S_n} \times [\epsilon_n \rho^{\prime}_n + (1 - \epsilon_n) \rho_n]\right ).
\label{eqnA4}
\end{equation}
The last term in this equation accounts for the different reflectance values of both treated and untreated surfaces.

The expected change in radiance in a DNB pixel that includes at least part of a neighborhood receiving CPT is
\begin{align}
{\Delta}L_n &= L_{n}(2) - L_{n}(1)\\
		    &=\frac{\Phi_n}{{\pi}S_n} \times [\epsilon_n \rho^{\prime}_n + (1 - \epsilon_n) \rho_n - \rho_n]\\
		    &=\frac{\epsilon_n \Phi_n (\rho^{\prime}_n - \rho_n)}{{\pi}S_n},
\end{align}
and the relative change in radiance due to the application of CPT materials is
\begin{equation}
\frac{L_{n}(2) - L_{n}(1)}{L_{n}(1)} = \frac{{\Delta}L_n}{L_{n}(1)} = \frac{\frac{\epsilon_n \Phi_n (\rho^{\prime}_n - \rho_n)}{{\pi}S_n}}{L_{n,bg} + L_{n,pd} + \frac{\rho_n}{\pi}\frac{\Phi_n}{S_n}}.
\label{eqnA8}
\end{equation}
Given the type of land use in the treated neighborhoods (namely low-density, single-story residential buildings), and because the main illumination source in these areas is fully shielded roadway luminaires, very little light is emitted in the sideways direction. We therefore expect that the greatest single contribution to the measured DNB radiances is attributable to ground reflections from streetlights. We furthermore assume that because of the urban setting of the DNB pixels in question the contributions of $L_{n,bg}$ and $L_{n,pd}$ are small compared to $L_{n,pr}$ and may therefore be dropped from Equation~\ref{eqnA8}. It then becomes
\begin{equation}
\frac{{\Delta}L_n}{L_{n}(1)} \approx \frac{\epsilon_n(\rho^{\prime}_n - \rho_n)}{\rho_n} = \epsilon_n \left( \frac{\rho^{\prime}_n}{\rho_n}-1 \right).
\label{eqnA9}
\end{equation}

To this point the result obtained (Equation~\ref{eqnA9}) is for a single DNB pixel. It can be extended to an arbitrary array of $N$ contiguous pixels if all individual pixels are of equal spatial area. The radiance of such an array is therefore the average value of the individual radiances of pixels in the array. The equivalent of Equation~\ref{eqnA3} in this situation is:
\begin{equation}
L(1) = \frac{1}{N} \sum_{n=1}^{N} L_{n}(1) = \frac{1}{N}\sum_{n=1}^{N}\frac{\rho_n}{\pi}\frac{\Phi_n}{S_n}.
\end{equation}
If (1) the distributions of irradiances and reflectances are not correlated; (2) the average reflectance of surfaces contained in a given DNB pixel is about the same as that in all other individual pixels (that is, $\rho_n\approx\rho$); and (3) the pixels are all of equal spatial area ($S_{n} = S_{p}$), then
\begin{equation}
L(1) = \frac{\rho}{{\pi}N}\sum_{n=1}^{N}\frac{\Phi_n}{S_n} = \frac{\rho}{{\pi}NS_{p}}\sum_{n=1}^{N}\Phi_n=\frac{\rho\Phi}{{\pi}S},
\end{equation}
where $\Phi=\sum_{n=1}^{N}\Phi_{n}$ is the total flux emitted in the DNB pixel array and $S=NS_{p}$ is the spatial area of the pixel array. The equivalent expression for $L_{n}(2)$ is
\begin{align}
L(2) &= \frac{1}{N} \sum_{n=1}^{N} L_{n}(2)\\
	     &= \frac{1}{N} \sum_{n=1}^{N} \frac{\Phi_n}{{\pi}S_n} \times [\epsilon_n \rho^{\prime}_n + (1 - \epsilon_n) \rho_n].
\end{align}
If reflectances before and after CPT materials are applied to streets in treated neighborhoods are similar across the individual DNB pixels (i.e., $\rho_n\approx\rho$ and $\rho^{\prime}_n\approx\rho^{\prime}$) and the individual DNB pixel areas are equal ($S_{n}=S_{p}$), then as before:
\begin{equation}
L(2) = \frac{1}{{\pi}NS_{p}}\sum_{n=1}^{N} [\epsilon_n \rho^{\prime} + (1 - \epsilon_n) \rho] \Phi_n.
\end{equation}
The expected change in radiance due to applying CPT is therefore:
\begin{align}
{\Delta}L &= L(2) - L(1)\\
		  &=\frac{1}{{\pi}S_{p}}\sum_{n=1}^{N} [\epsilon_n \rho^{\prime} + (1 - \epsilon_n) \rho] \Phi_n - \rho\Phi_{n}\\
		  &=\frac{\rho^{\prime}-\rho}{{\pi}S}\sum_{n=1}^{N}\Phi_{n}\epsilon_{n}.
\end{align}
Since land uses in all the DNB pixels are about the same, their light emissions are expected to be very similar. That is, we do not expect especially large radiance variations from one DNB pixel to the next, and therefore $\Phi_{n}=\Phi/N$. The expected radiance change then reduces to
\begin{equation}
{\Delta}L_n = \frac{\rho^{\prime}-\rho}{{\pi}S}\sum_{n=1}^{N}\frac{\Phi}{N}\epsilon_{n}=\frac{\rho^{\prime}-\rho}{{\pi}S}\frac{\Phi}{N}\sum_{n=1}^{N}\epsilon_{n}=\frac{\rho^{\prime}-\rho}{{\pi}S}\Phi\epsilon,
\end{equation}
in particular because
\begin{equation}
\frac{1}{N}\sum_{n=1}^{N}\epsilon_{n}=\frac{1}{NS_{p}}\sum_{n=1}^{N}\epsilon_{n}S_p=\frac{1}{S}\sum_{n=1}^{N}\epsilon_{n}S_{p}=\frac{\epsilon}{N}
\end{equation}
and $S=NS_{p}$. Then the relative change becomes
\begin{equation}
\frac{{\Delta}L}{L(1)}=\frac{\frac{\rho^{\prime}-\rho}{{\pi}S}\Phi\epsilon}{\frac{\rho\Phi}{{\pi}S}}=\epsilon\left(\frac{\rho^{\prime}}{\rho}-1\right).
\end{equation}
The expected change in radiance in an array of $N$ contiguous DNB pixels therefore depends only on the fraction $\epsilon$ of the spatial area of the pixels to which CPT materials are applied and the reflectivity of the surfaces they cover before and after the treatment.

%
%
\section*{Acknowledgements}
\label{sec:ack}
The author wishes to thank Prof.~Salvador Bar\'{a} for advice and useful discussions in particular regarding the modeling work described here.

%
%
\bibliographystyle{elsarticle-num} 
\bibliography{barentine-cool-pavements}

\end{document}